\language1
\mag=\magstep1
\def\nb #1{{\hbox{\bf #1}}}
\font\bsym=cmbsy10
\def\bnabla{\hbox{$\textfont2=\bsym \nabla$}}
\count0=1
\font\huge=cmr12
\font\biggb=cmbx12 scaled \magstep2
\font\ini=cmbx8 scaled \magstep1
\hsize=16truecm
\vsize=23truecm
\tolerance=10000
\hbadness=10000
\hfuzz=20pt
\footline={\ifnum\pageno=1
              \relax
           \else
\ifodd\pageno\rightfootline
              \else\leftfootline
              \fi
           \fi}
\def\leftfootline{\hfill{\bf\folio}\hfill
}
\def\rightfootline{\hfill
 \bf{\folio}\hfill}

\vbox{\vskip 1.5truecm}
\centerline{\biggb Vortex dynamics in 2D antiferromagnets}

\vbox{\vskip 2truecm}

\centerline{\huge S. Komineas and N. Papanicolaou}
\centerline{ Department of Physics, University of Crete,}
\centerline{ and Research Center of Crete,}
\centerline{ Heraklion, Greece}

\vbox{\vskip 6truecm}

\centerline{\bf Abstract}
\bigskip
{ The dynamics of vortices in a 2D Heisenberg antiferromagnet with
an easy-plane anisotropy is studied numerically within the discrete spin
model as well as analytically within a continuum approximation based on a
suitable extension of the relativistic nonlinear $\sigma$ model. We find that
two like vortices scatter at $90^{\circ}$ during a head-on collision,
whereas a vortex-antivortex pair is annihilated into spinwave radiation
emitted mainly at $90^{\circ}$. When a uniform bias field is applied,
vortex dynamics is affected rather profoundly and acquires the characteristic
features of the Hall effect of electrodynamics or the Magnus effect
of fluid dynamics. In particular, a single vortex is always spontaneously 
pinned, two like vortices form a rotating bound state, and a 
vortex-antivortex pair undergoes Kelvin motion. Finally, in the presence
of a bias field, vortices are shown to be the prominent topological
excitations even for an isotropic antiferromagnet.}

\vfill
\eject

\bigskip
\centerline{\ini I.\ INTRODUCTION}
\bigskip
Topological magnetic solitons have been studied extensively in the case of
ferromagnets (FM) and weak ferromagnets (WFM). In both cases a nonvanishing
 magnetization develops in the ground state, albeit by a different 
physical mechanism, which allows a detailed experimental investigation by
standard techniques [1,2]. In contrast, direct experimental evidence
for pure antiferromagnetic (AFM) solitons is limited. Nevertheless,
theoretical arguments suggest that static AFM solitons should exist
for essentially the same reason as in ordinary ferromagnets, even though
their dynamics is significantly different.

The dynamics is now governed by suitable extensions of the relativistic
nonlinear $\sigma$ model [2] instead of the Landau-Lifshitz equation [1].
Therefore dynamical concepts familiar from the theory of FM domain walls
and bubbles need to be reanalyzed within the relativistic theory. It is the
general purpose of this paper to pursue a study of the dynamics of
two-dimensional (2D) AFM solitons that emphasizes the influence of the
underlying topological structure. Our starting point is some recent work
on AFM and WFM domain walls [3] which revealed that some important issues
in the derivation of the associated nonlinear $\sigma$ model had been
mistreated in earlier treatments. We were thus sufficiently motivated
to extend the analysis of Refs. [3] to the case of layered or 2D
antiferromagnets whose significance has increased in recent years in
connection with high-$T_c$ superconductivity.

The present work proceeds by a combination of numerical and
analytical methods. Numerical calculations were performed within the 
standard discrete spin model. However a transparent interpretation of the
numerical results could not be achieved without the aid of a continuum
approximation at the heart of which lies the relativistic nonlinear
$\sigma$ model. A complete account of the continuum model is given in
Sec. II where we discuss, in particular, the relevance of certain 
parity-breaking contributions that are implicit in spin models involving
antiferromagnetic interactions. Numerical and analytical results are then
combined in Sec. III to provide a complete description of static AFM 
vortices.

Subsequent sections are devoted to a detailed study of some special dynamical
features due to the underlying topology. For instance, head-on collisions
of vortices are examined in Sec. IV and shown to exhibit a characteristic
$90^{\circ}$ scattering pattern familiar from similar studies of relativistic
skyrmions and monopoles [4]. The most surprising element is described
in Sec. V where it is shown that an applied uniform magnetic field affects
vortex dynamics rather profoundly. A direct link between topology and
dynamics is established by means of the conservation laws of linear and angular
momentum expressed as moments of a suitable topological vorticity, in analogy
with related work for magnetic bubbles in ordinary ferromagnets [5,$\!$ 6].
Possible phenomenological implications of the derived dynamical picture
are further discussed in the concluding Sec.$\!$ VI. Finally, an Appendix
presents some useful virial relations that generalize the 
well-known scaling theorem of Derrick [7].
\vfill
\eject

\bigskip
\bigskip
\centerline{\ini II.\ THE NONLINEAR $\sigma$ MODEL}
\bigskip
We shall study a spin system described by the Hamiltonian
$$W=\sum_{ij}[J\,\nb S_{i, j}\cdot (\nb S_{i+1, j}+\nb S_{i, j+1})
+{1\over 2}g(S_{i, j}^3)^2],\eqno (2.1)$$
where the summation extends over all sites of a square lattice,
$\nb S_{i, j}$ is the spin vector at site $(i, j)$ and $S_{i, j}^3$
its third component. Both the exchange constant $J$ and the anisotropy
constant $g$ are taken to be positive and hence (2.1) describes a 
Heisenberg antiferromagnet with an easy-plane single ion anisotropy.

The spin variables are treated as classical vectors with constant
magnitude $s$ and satisfy the equation of motion
$${\partial\nb S_{i, j}\over\partial t}=\nb S_{i, j}\times\nb F_{i, j},\qquad
\nb S_{i, j}^2=s^2,\eqno (2.2)$$
where the effective field $\nb F$ is determined from the general relation
$$\nb F_{i, j}=-{\partial W\over\partial\nb S_{i, j}},\eqno (2.3)$$
or, more explicitly, by
$$\nb F_{i, j}=-J(\nb S_{i+1, j}+\nb S_{i-i, j}+\nb S_{i, j+1}+
\nb S_{i, j-1})-gS^3_{i, j}\nb e,\eqno (2.4)$$
where $\nb e=(0, 0, 1)$ is a unit vector along the third direction.
The exchange contribution in the effective field (2.4) contains fewer
than four terms on the perimeter of a finite lattice, the precise number
of such terms being equal to the number of nearest neighbors. All
dynamical simulations presented in this paper will be based on the above
relatively simple set of discrete equations adapted to an open finite lattice.

However, in order to pursue an efficient study of the dynamics, we also
sonsider a continuum approximation which is possible at weak anisotropy,
$$\varepsilon = \sqrt{{g\over J}}\ll 1,\eqno (2.5)$$

\noindent
where soliton structures extend over a large number of sites given roughly
by $1/\varepsilon$. The continuum limit is not completely straightforward
because of the implicit antiferromagnetic discontinuity as one moves
from site to site. It is then important to first identify dynamical 
variables that may possess a smooth limit as $\varepsilon\to 0$.

In one dimension continuity is achieved by a simple dimerization process [3].
A square lattice may also be thought of as a collection of dimers, as 
illustrated in Fig.$\!$ 1 for a finite lattice cut along the diagonals.
The sites of the original square lattice are depicted by two sets of 
(solid and open) circles which form two intertwining sublattices. This
designation is used merely to indicate that the two spins of any given
dimer belong to different sublattices. In particular, the ground 
(N\'{e}el) state is such that the two spins point in opposite directions 
but are uniform on each sublattice. Because of the easy-plane anisotropy
the N\'{e}el state is polarized along any direction in the plane perpendicular
to the third axis. This azimuthal degeneracy of the ground state will play
an important role in the following.

A generic dimer $AB$ (see Fig.$\!$ 1) is labeled by a pair of indices
$(\alpha , \beta)$ numbered consecutively 
$(\alpha , \beta =1, 2, \ldots , N)$. In our explicit illustrations $N$
is taken to be even, but this is only a minor technical assumption.
Let us denote by $\nb A_{\alpha , \beta}$ and $\nb B_{\alpha , \beta}$
the two spins of dimer $AB$. Eq. (2.2) is then written as a system of two
coupled equations,
$${\partial\nb A_{\alpha , \beta}\over\partial t}=
\nb A_{\alpha , \beta}\times\nb F_{\alpha , \beta}\qquad
{\partial\nb B_{\alpha , \beta}\over\partial t}=
\nb B_{\alpha , \beta}\times\nb G_{\alpha , \beta},\eqno (2.6)$$
where the effective fields $\nb F$ and $\nb G$ are given by
$$\eqalign{\nb F_{\alpha , \beta} & = -J(\nb B_{\alpha , \beta}+
\nb B_{\alpha -1, \beta}+\nb B_{\alpha , \beta -1}+
\nb B_{\alpha -1, \beta -1})-gA_{\alpha , \beta}^3\nb e,\cr
\noalign{\medskip}
\nb G_{\alpha , \beta} & = -J(\nb A_{\alpha , \beta}+
\nb A_{\alpha +1, \beta}+\nb A_{\alpha , \beta +1}+
\nb A_{\alpha +1, \beta +1})-gB_{\alpha , \beta}^3\nb e,\cr}\eqno (2.7)$$
for a generic point inside the lattice. The exchange contributions
in Eq. (2.7) contain only two terms for each point on the perimeter of the 
finite lattice of Fig. 1.

As a first step in the derivation of a continuum approximation we 
introduce the discrete set of variables
$$\eta =\sqrt{2}\, \varepsilon (\alpha -\alpha_0),\qquad
\xi =\sqrt{2}\, \varepsilon (\beta -\beta_0),\eqno (2.8)$$
which become continuous in the limit $\varepsilon\to 0$ and provide a 
measure of distances along the diagonals of the square lattice. Actual
distances are given by $a\eta/\varepsilon$ and $a\xi/\varepsilon$ where $a$
is the physical distance between two neighboring magnetic ions. The lattice
constant $a$ will not appear in our theoretical development except
when various quantities of interest will have to be translated in
physical units. Finally the convenient choice 
$\alpha_0=(N+1)/2=\beta_0$ in Eq. (2.8) sets the origin of the coordinate
system at the center of the lattice of Fig. 1.

The main assumption supported by numerical calculations is that the spin
variables $\nb A_{\alpha , \beta}$ and $\nb B_{\alpha , \beta}$ approach
smooth continuum limits $\nb A=\nb A(\eta, \xi)$ and 
$\nb B=\nb B(\eta, \xi)$ at weak anisotropy $(\varepsilon\to 0)$. Then
we make the substitutions $\nb A_{\alpha , \beta}\to\nb A$ and 
$\nb B_{\alpha , \beta}\to \nb B$ in Eqs.$\!$ (2.6)-(2.7) together with the
Taylor expansions 
$$\eqalign{\nb A_{\alpha\pm 1, \beta} & \to \nb A\pm\delta\nb A_{\eta}
+{1\over 2}\delta^2\nb A_{\eta\eta},\cr
\noalign{\medskip}
\nb A_{\alpha , \beta \pm 1} & \to \nb A\pm\delta\nb A_{\xi}
+{1\over 2}\delta^2\nb A_{\xi\xi},\cr
\noalign{\medskip}
\nb A_{\alpha\pm 1, \beta \pm 1} & \to \nb A\pm\delta
(\nb A_{\eta}+\nb A_{\xi})+{1\over 2}\delta^2
(\nb A_{\eta\eta}+\nb A_{\xi\xi}+2\nb A_{\eta\xi}),\cr}\eqno (2.9)$$
and similar expansions for the field $\nb B$. Here subscripts denote
differentiation with respect to the indicated arguments and
$$\delta =\sqrt{2}\, \varepsilon\eqno (2.10)$$
is used as a temporary notational abbreviation. Eqs. (2.6) are then
approximated by
$${\partial\nb A\over\partial t}=\nb A\times\nb F,\qquad
{\partial\nb B\over\partial t}=\nb B\times\nb G,\eqno (2.11)$$
where
$$\eqalign{\nb F & = -J[4\nb B-2\delta (\nb B_{\eta}+\nb B_{\xi})+\delta^2
(\nb B_{\eta\eta}+\nb B_{\xi\xi}+\nb B_{\eta\xi})]-gA_3\nb e,\cr
\noalign{\medskip}
\nb G & = -J[4\nb A+2\delta (\nb A_{\eta}+\nb A_{\xi})+\delta^2
(\nb A_{\eta\eta}+\nb A_{\xi\xi}+\nb A_{\eta\xi})]-g B_3\nb e,\cr}
\eqno (2.12)$$
This system of equations is not yet fully consistent because it appears to mix
different powers of $\delta$ (or $\varepsilon$).

In order to obtain a consistent continuum model we proceed as in the 1D
case studied in Ref. [3]. First, we introduce the linear combination
of fields
$$\nb m={1\over 2s}(\nb A+\nb B),\qquad
\nb n={1\over 2s}(\nb A-\nb B),\eqno (2.13)$$
which satisfy the constraints
$$\nb m\cdot\nb n=0,\qquad \nb m^2+\nb n^2=1.\eqno (2.14)$$
Second, we define a dimensionless time variable
$$\tau =2\delta sJt=2\sqrt{2}\, \varepsilon sJt.\eqno (2.15)$$
An equivalent form of Eqs. (2.11) then reads
$$\eqalign{\delta{\partial\nb m\over\partial\tau} = & - \delta
[(\nb m\times\nb n)_{\eta}+(\nb m\times\nb n)_{\xi}]+{1\over 2}\delta^2
[\nb n\times (\nb n_{\eta\eta}+\nb n_{\xi\xi}+\nb n_{\eta\xi})\cr
\noalign{\medskip}
& - \nb m\times (\nb m_{\eta\eta}+\nb m_{\xi\xi}+\nb m_{\eta\xi})]
-{1\over 4}\delta^2[m_3(\nb m\times\nb e)+n_3(\nb n\times\nb e)],\cr
\noalign{\bigskip}
\delta{\partial\nb n\over\partial\tau} = &\, 4(\nb m\times\nb n)+\delta
[\nb m\times (\nb m_{\eta}+\nb m_{\xi})-\nb n\times
(\nb n_{\eta}+\nb n_{\xi})]\cr
\noalign{\medskip}
& +{1\over 2}\delta^2[\nb m\times (\nb n_{\eta\eta}+\nb n_{\xi\xi}
+\nb n_{\eta\xi})-
\nb n\times (\nb m_{\eta\eta}+\nb m_{\xi\xi}+\nb m_{\eta\xi})]\cr
\noalign{\medskip}
& - {1\over 4}\delta^2[m_3 (\nb n\times\nb e)+n_3
(\nb m\times\nb e)].\cr}\eqno (2.16)$$
A simple inspection of the above equations suggests that consistency
is obtained if $\nb m$ is of order $\delta$. Then the leading 
approximation of the second equation is 
$$\delta{\partial \nb n\over\partial\tau}=4(\nb m\times\nb n)-\delta
[\nb n\times (\nb n_{\eta}+\nb n_{\xi})]\eqno (2.17)$$
and the constraints of Eq. (2.14) reduce to
$$\nb m\cdot\nb n=0,\qquad \nb n^2=1,\eqno (2.18)$$
to within terms of order $\delta^2$. Therefore taking the cross product
of both sides of Eq. (2.17) with $\nb n$ and using the constraints (2.18)
yields
$$\nb m={\varepsilon\over 2\sqrt{2}}[-(\nb n_{\eta}+\nb n_{\xi})+
(\nb n\times\dot{\nb n})],\eqno (2.19)$$
where we have restored the small parameter $\varepsilon$ from Eq. (2.10).
Finally Eq. (2.19) is inserted in the first of Eqs. (2.16) to give in the 
limit $\varepsilon\to 0$ the differential equation
$$\nb n\times\nb f=0,\qquad \nb f=\ddot{\nb n}-\Delta\nb n+n_3\nb e,
\eqno (2.20)$$
where $\Delta$ is the 2D Laplacian
$$\Delta ={\partial^2\over\partial\eta^2}+{\partial^2\over\partial\xi^2}.
\eqno (2.21)$$

\noindent
It is understood that terms of order $\varepsilon^2$ in Eqs. (2.18), (2.20)
and order $\varepsilon^3$ in Eq. (2.19) have been neglected.

Therefore the continuum approximation is governed mainly by Eq. (2.20) which
is a simple extension of the relativistic nonlinear $\sigma$ model to
include a single ion anisotropy. The corresponding velocity of light
is equal to unity thanks to our choice of rationalized space and time
variables. In our conventions, the spin magnitude $s$ is a simple multiple
of the Planck constant (e.g., $s={1\over 2}\hbar$) and thus carries
dimension of action, $sJ$ of frequency and $s^2J$ of energy. The ratio 
$\varepsilon$ of Eq. (2.5) is dimensionless, so is the time variable 
$\tau$ of Eq. (2.15). Also recalling that actual distances are given by
$a\eta/\varepsilon$ and $a\xi/\varepsilon$, where $a$ is the lattice
constant, we conclude that velocity is measured in units of
$$c=2\sqrt{2}asJ,\eqno (2.22)$$
which coincides with the phase velocity of pure AFM magnons on a square
lattice in the long-wavelength limit. More generally, the magnon velocity
is given by
$$c=2asJ\sqrt{{z\over 2}},\eqno (2.23)$$
where $z$ is the lattice coordination number. Applied for a square lattice
$(z=4)$ Eq.$\!$ (2.23) reduces to Eq. (2.22), whereas for a chain $(z=2)$
it yields the limiting velocity $c=2asJ$ of Refs. [3].

Needless to say, a complete description of the original spin model requires
knowledge of both fields $\nb m$ and $\nb n$. On the other hand, given
a solution $\nb n=\nb n(\eta , \xi , \tau)$ of the nonlinear $\sigma$ model,
the field $\nb m$ may be determined from Eq.$\!$ (2.19) by simple 
differentiations and may thus be viewed as an auxiliary field. Yet this
apparently straightforward result has been controversial. For instance,
the parity-breaking gradient terms in Eq.$\!$ (2.19) 
were recently derived within the 1D model and shown
to be important for various structural properties of AFM and WFM domain
walls [3]. The possible occurrence of such terms had been anticipated
on symmetry grounds [8] but this possibility was overlooked in the 
literature for a long time [9].

One should stress that symmetry arguments do not predict the precise
coefficients of the parity-breaking gradient terms in Eq. (2.19) and are
generally susceptible to overinterpretation. The danger from this 
dimerization ambiguity is already present in one dimension and was completely
analyzed in Refs. [3]. The situation is only compounded in two dimensions.
For example, had we chosen to work with a horizontal instead of the vertical
dimerization of Fig. 1, the nonlinear $\sigma$ model of Eq.$\!$ (2.20) would 
not be affected but the gradient terms in Eq. (2.19) would appear as
$\nb n_{\eta}-\nb n_{\xi}$ instead of $\nb n_{\eta}+\nb n_{\xi}$. This is an
indication that the local values of the field $\nb m$ are sensitive to the
mode of dimerization and cannot be literally interpreted as magnetization.
However no real mathematical or physical ambiguity appears when the results
of a calculation or an experiment are consistently interpreted in reference
to a specific mode of dimerization. But an attempt to measure the
``magnetization'' $\nb m$ by standard techniques may lead to a fuzzy 
magnetization curve around any nontrivial soliton structure, unless spin
values can be resolved at every site [3].

To press the above picture further we note that dimerization is not the
only way to achieve continuity on a square antiferromagnetic lattice.
We may also consider the tetramerous configuration illustrated for a finite
lattice in Fig. 2. Each tetramer is again labeled by two indices $\alpha$
and $\beta$ numbered consecutively $(\alpha , \beta = 1, 2, \ldots , N)$
and spin values are smooth as one moves from corresponding sites of
one tetramer to the next. Let us again denote by $\nb A_{\alpha , \beta}$,
$\nb B_{\alpha , \beta}$, $\nb C_{\alpha , \beta}$ and
$\nb D_{\alpha , \beta}$ the spins on a generic tetramer $ABCD$ shown in
Fig.$\!$ 2, in terms of which the original equation of motion (2.2) may be
written as a system of four coupled equations analogous to Eqs.$\!$ (2.6).
This new system may then be used for the derivation of the continuum
approximation by a method similar to the one explained earlier within
the dimerization scheme of Fig. 1.

We omit the lengthy algebraic details and simply state the final results.
We now use the Cartesian coordinates
$$x=2\varepsilon (\alpha -\alpha_0),\qquad y=2\varepsilon (\beta -\beta_0),
\eqno (2.24)$$
which are related to the coordinates $\eta$ and $\xi$ of Eq. (2.8) by a
$45^{\circ}$ rotation. Here the origin of the coordinate system is
again set at the center of the lattice of Fig. 2 by choosing the arbitrary
constants as $\alpha_0=(N+1)/2=\beta_0$. Next we assume that the four spins
on a tetramer approach smooth continuum limits when $\varepsilon\to 0$ 
denoted by $\nb A, \nb B, \nb C$ and $\nb D$ which are some functions of 
the spatial coordinates $x$ and $y$ of Eq. (2.24) and the time variable 
$\tau$ of Eq. (2.15). A more convenient set of fields is given by
$$\eqalign{\nb m & = {1\over 4s}(\nb A+\nb B+\nb C+\nb D),\cr
\noalign{\medskip}
\nb k & = {1\over 4s}(\nb A+\nb B-\nb C-\nb D),\cr}\qquad
\eqalign{\nb n & = {1\over 4s}(\nb A-\nb B+\nb C-\nb D),\cr
\noalign{\medskip}
\nb l & = {1\over 4s}(\nb A-\nb B-\nb C+\nb D),\cr}\eqno (2.25)$$
and satisfy the constraints
$$\eqalign{\nb m^2+\nb n^2+\nb k^2+\nb l^2 & = 1,\cr
\noalign{\medskip}
\nb m\cdot\nb n+\nb k\cdot\nb l & = 0,\cr}\qquad
\eqalign{\nb m\cdot\nb k+\nb n\cdot\nb l & = 0,\cr
\noalign{\medskip}
\nb m\cdot\nb l+\nb n\cdot\nb k & = 0.\cr}\eqno (2.26)$$

In the strict continuum limit $(\varepsilon\to 0)$ the constraints
reduce to
$$\nb m\cdot\nb n=\nb k\cdot\nb n=\nb l\cdot\nb n=0,\qquad
\nb n^2=1,\eqno (2.27)$$
the fields $\nb m, \nb k$ and $\nb l$ are expressed in terms of $\nb n$ by
$$\nb m={\varepsilon\over2\sqrt{2}}(\nb n\times\dot{\nb n}),\qquad
\nb k=-{\varepsilon\over2}\nb n_x,\qquad
\nb l=-{\varepsilon\over2}\nb n_y,\eqno (2.28)$$
and $\nb n$ itself satisfies the differential equation
$$\nb n\times\nb f=0,\qquad 
\nb f=\ddot{\nb n}-\Delta\nb n+n_3\nb e,\eqno (2.29)$$
where $\Delta$ is the 2D Laplacian
$$\Delta ={\partial^2\over\partial x^2}+{\partial^2\over\partial y^2}.
\eqno (2.30)$$
It is again understood that Eqs. (2.27)-(2.29) are accurate to within
terms of order $\varepsilon^2$.

Although the field $\nb n$ defined in Eq.$\!$ (2.25) is not directly related
to the field $\nb n$ of Eq. (2.13), its continuum dynamics is still
governed by the nonlinear $\sigma$ model of Eq.$\!$ (2.29) which is 
equivalent to Eq. (2.20). However a clear distinction between the two
formulations emerges at the level of the auxiliary fields. We now need
three such fields ($\nb m, \nb k$ and $\nb l$) which are expressed in terms
of $\nb n$ through Eqs.$\!$ (2.28). These relations reenforce our earlier
remarks concerning parity-breaking contributions, namely that they are
sensitive to the specific mode of taking the continuum limit. For 
instance, such contributions are no longer present in the field $\nb m$
but their effect is accounted for by the new auxiliary fields $\nb k$
and $\nb l$. Once again we must conclude that the local values of the 
field $\nb m$ cannot be interpreted literally as magnetization.

The perplexing nature of the auxiliary fields calls for a summary of our
main strategy. The continuum model is extensively used to motivate the
results discussed in subsequent sections. But all numerical calculations
are based on the original discrete equation (2.2). The actual 
calculations may be performed on an open finite lattice either of the type
shown in Fig. 1 or that of Fig. 2. The explicit results may then be 
employed to construct the fields $\nb m$ and $\nb n$ in the former case
and the fields $\nb m, \nb n, \nb k$ and $\nb l$ in the latter. In both
cases the validity of the respective continuum models can be verified
explicitly for a sufficiently weak anisotropy $(\varepsilon\ll 1)$.
Nonetheless several apparent paradoxes emerged in the course of our
investigation which were all resolved in favor of the formulation
presented in this section.

\bigskip
\bigskip
\centerline{\ini III.\ STATIC VORTICES}
\bigskip
We consider first the problem of finding interesting static solutions within
the continuum model. In both formulations developed in the previous section,
the basic issue is to determine the field $\nb n$ from the nonlinear
$\sigma$ model and then proceed with the calculation of the auxiliary fields.
Time derivatives vanish in the static limit and one obtains the reduced
equations
$$\nb n\times\nb f=0,\qquad \nb f=-\Delta\nb n+n_3\nb e.\eqno (3.1)$$
It proves convenient to derive the field $\nb f$ from a variational argument,
$$\nb f={\delta W\over\delta\nb n},\qquad
W={1\over 2}\int [(\partial_{\mu}\nb n\cdot\partial_{\mu}\nb n)
+n^2_3]dxdy,\eqno (3.2)$$
where $W$ is the energy functional in the static limit (see Sec.$\!$ V).
The repeated index is summed over two distinct values $\mu =1$ and 2
corresponding to the two spatial coordinates $x$ and $y$ (or $\eta$
and $\xi$). It is also useful to resolve the constraint $\nb n^2=1$
explicitly using, for example, the spherical parametrization
$$n_1=\sin\Theta\cos\Phi ,\qquad n_2=\sin\Theta\sin\Phi ,\qquad
n_3=\cos\Theta\eqno (3.3)$$
in terms of which
$$W={1\over 2}\int [(\partial_{\mu}\Theta\partial_{\mu}\Theta)+
\sin^2\Theta\, (\partial_{\mu}\Phi\partial_{\mu}\Phi)+\cos^2\Theta]dxdy
\eqno (3.4)$$
and Eqs. (3.1) are equivalent to
$${\delta W\over\delta\Theta}=0={\delta W\over\delta\Phi},\eqno (3.5)$$
or
$$\Delta\Theta +[1-(\partial_{\mu}\Phi\partial_{\mu}\Phi)]\cos\Theta
\sin\Theta=0,\qquad
\partial_{\mu}(\sin^2\Theta\, \partial_{\mu}\Phi)=0.\eqno (3.6)$$
It should be noted that the above static equations are formally identical
to those encountered in easy-plane ferromagnets. However the physical
interpretation of the field $\nb n$ is now different and the actual 
construction of the corresponding AFM solitons on the discrete lattice is
effected by the prescriptions of Sec. II.

The simplest solution of Eqs. (3.6) is the ground state configuration
$\Theta ={\pi\over 2}$ and $\Phi =\phi_0$, where $\phi_0$ is an
arbitrary constant, for which the energy achieves its absolute minimum
$(W=0)$. Such a simple configuration $(n_1=\cos\phi_0$, $n_2=\sin\phi_0$,
$n_3=0)$ is consistent with our earlier remark that the ground state is the
usual N\'{e}el state with spins polarized along any direction in the (12)
plane. Nontrivial static solutions are also stationary points of the 
energy functional $W$ and are subject to limitations imposed by the
familiar scaling theorem of Derrick [7]. Applied for the functional (3.4)
Derrick's argument yields the virial relation
$$\int\cos^2\Theta\,  dx dy=0.\eqno (3.7)$$
Therefore one must conclude that nontrivial static solutions with finite
energy do not exist.

The above argument does not exclude interesting solutions with infinite
energy such as vortices. In fact, the possible existence of vortices is
probed by a suitable generalization of the Derrick theorem worked out
in the Appendix. Eq. (3.7) is then replaced by
$$\int\cos^2\Theta\, dx dy=\pi ,\eqno (3.8)$$
which not only does not exclude the possibility of vortex solutions but
predicts the actual value of their anisotropy energy. However the 
exchange energy is logarithmically divergent.

To obtain an explicit solution we consider the usual cylindrical
coordinates $x=\rho\cos\phi$, $y=\rho\sin\phi$ and make the following
ansatz for a vortex located at the origin:
$$\Theta =\theta (\rho),\qquad \Phi =\kappa (\phi +\phi_0),\eqno (3.9)$$
where $\kappa =\pm1$ will be referred to as the vortex number and
$\phi_0$ is an arbitrary constant reflecting the azimuthal symmetry.
Then the second equation in (3.6) is automatically satisfied and the first
yields
$${1\over\rho}{\partial\over\partial\rho}\left(\rho
{\partial\theta\over\partial\rho}\right) +
\left( 1-{1\over\rho^2}\right)\cos\theta\sin\theta =0.\eqno (3.10)$$
The energy (3.4) is accordingly reduced to
$$W={1\over 2}\int^{\infty}_0
\left[\left({\partial\theta\over\partial\rho}\right)^2+
{\sin^2\theta\over\rho^2}+\cos^2\theta\right] (2\pi\rho d\rho)\eqno (3.11)$$
and its stationary points are solutions of Eq.$\!$ (3.10). This equation is
consistent with the asymptotic behavior
$$\theta\mathop{\approx}\limits_{\rho\to 0}c_1\rho ,\qquad
\theta\mathop{\approx}\limits_{\rho\to\infty}{\pi\over 2}-
{c_2e^{-\rho}\over\sqrt{\rho}},\eqno (3.12)$$
where $c_1$ and $c_2$ are some constants. The actual solution depicted in
Fig.$\!$ 3 was calculated numerically via a relaxation algorithm applied to the
restricted energy functional (3.11) taking into account the boundary
values (3.12).

Furthermore we note that for every solution $\theta$ of Eq. (3.10)
a new solution is obtained by the formal replacement
$\theta\to\pi -\theta$. Hence the complete result for the field
$\nb n=(n_1, n_2, n_3)$ reads
$$n_1=\sin\theta\cos [\kappa (\phi +\phi_0)],\qquad
n_2=\sin\theta\sin [\kappa (\phi +\phi_0)],\qquad
n_3=\nu\cos\theta,\eqno (3.13)$$
where $\theta =\theta (\rho)$ is taken from Fig. 3, while the vortex number
$\kappa$ and the polarity $\nu$ are given by
$$\kappa =\pm 1, \qquad \nu =\pm 1,\eqno (3.14)$$
taken in any combination. In other words, both the vortex $(\kappa =1)$
and the antivortex $(\kappa =-1)$ come in two varieties
$(\nu =\pm 1)$.

The asymptotic behavior (3.12) makes it evident that the centrifugal
(second) term in the energy (3.11) diverges logarithmically with the
size of the system but the remaining terms are finite. In fact, the
numerically calculated anisotropy contribution was found to be in
excellent agreement with virial relation (3.8). The weak (logarithmic)
divergence of the total energy is not necessarily an obstacle to the
production of vortices in an antiferromagnet. In this respect, one should
recall that vortices are easily produced in a rotating cylindrical bucket
filled with an ordinary fluid or a superfluid [10]. Furthermore 
vortex-antivortex pairs have finite energy and should play an important
role in thermodynamics.

Keeping with the strategy outlined in the concluding paragraph of Sec.$\!$ II,
we must  comment on the manner in which Eq.$\!$ (3.13) furnishes a vortex on
the discrete lattice. The field $\nb n$ of Eq.$\!$ (3.13) may be used to 
calculate the auxiliary field $\nb m$ of Eq.$\!$ (2.19) or the fields
$\nb m, \nb k$ and $\nb l$ of Eq.$\!$ (2.28), applied for a static solution
$(\dot{\nb n}=0)$, and thereby determine the original spin configuration
on the lattices of Fig. 1 and Fig. 2, respectively. As a check of 
consistency we have used the resulting spin configuration as initial 
condition within a fully-dissipative algorithm [3] applied to the
complete energy functional (2.1) on a $200\times 200$ lattice. The
calculated relaxed state accurately reproduced both the profile of Fig. 3
and the corresponding auxiliary fields for a reasonably small value of the
anisotropy constant $(\varepsilon =0.1)$. The spin values on the discrete
lattice are partially illustrated in Fig. 4 for both a vortex and an
antivortex and exhibit the characteristic antiferromagnetic discontinuity
between the two sublattices. Finally we mention that a similar calculation
of an AFM vortex can be found in Ref. [11] using an Ising-like instead
of a single-ion anisotropy.

Since the vortex solutions described above are present for any finite
value of the anisotropy, however weak, it is also of interest to examine
the extreme limit of vanishing anisotropy. The static equations (3.1) then
reduce to
$$\nb n\times\Delta\nb n=0,\qquad \nb n^2=1\eqno (3.15)$$
and are equivalent to the O(3) nonlinear $\sigma$ model in a 2D 
Euclidean space. The small parameter $\varepsilon$ hidden in the definition
of spatial coordinates is no longer related to an anisotropy constant but
is an intrinsic scale set by the actual spread of localized solitons. The
implied arbitrariness of $\varepsilon$ is reflected by the scale invariance
of Eq.$\!$ (3.15) and is already an indication that the corresponding solitons
are metastable. Explicit solutions of Eq.$\!$ (3.15) are the well-known 
Belavin-Polyakov (BP) instantons [12] viewed here as static solitons in the
2+1 dimensional theory of current interest. They differ from the vortex
(3.13) mainly in the asymptotic value of the field $\nb n$ which is now 
given by
$$\nb n\mathop{\longrightarrow}\limits_{|\nb x|\to\infty}(0, 0, 1)
\eqno (3.16)$$ 
and coincides with the ground state configuration of the isotropic 
antiferromagnet defined up to a global O(3) rotation. Therefore the BP
instantons may be called AFM bubbles, by analogy to FM bubbles [1],
and are classified by the Pontryagin index or winding number $Q$ which is
defined as
$$Q={1\over 4\pi}\int q\,  dx dy,\qquad
q={1\over 2}\varepsilon_{\mu\nu}(\partial_{\nu}\nb n\times\partial_{\mu}\nb n)
\cdot\nb n,\eqno (3.17)$$
where $\varepsilon_{\mu\nu}$ is the 2D antisymmetric tensor, and is 
integer-valued $(Q=0,$ $\pm 1$, $\pm 2, \ldots)$. The topological density
$q$ may also be expressed in terms of the spherical variables as
$$q=\varepsilon_{\mu\nu}\sin\Theta\, \partial_{\nu}\Theta\partial_{\mu}
\Phi\eqno (3.18)$$
and plays an important role in the dynamical theory of FM bubbles [5,6].

An examination of the asymptotic behavior of Eq. (3.13) suggests that a
vortex may be viewed roughly as a half bubble. To push this remark further
we calculate the density $q$ for the vortex configuration (3.13),
$$q={\kappa\nu\over\rho}{\partial\cos\theta\over\partial\rho},\eqno (3.19)$$
where $\theta=\theta (\rho)$ must be taken from Fig. 3. Therefore
the winding number calculated from Eq. (3.17) is found to be 
$$Q={1\over 2}\kappa\nu [\cos\theta(\infty)-\cos\theta (0)]=-{1\over 2}
\kappa\nu\eqno (3.20)$$
and depends on both the vortex numper $\kappa$ and the polarity $\nu$.
This result confirms the vague notion that a vortex is topologically equivalent
to a half bubble $(Q=\pm {1\over 2})$. However a modified topological charge
that is related to the vortex number but not the polarity will arise
more naturally in the dynamical context of Sec. V.

\bigskip
\bigskip
\centerline{\ini IV.\ HEAD-ON COLLISIONS}
\bigskip
Although static AFM solitons are similar to those encountered in 
ferromagnets, their dynamics is significantly different. For comparison
purposes, it is useful to recall at this point the two main dynamical
features of FM bubbles: (a) An FM bubble cannot be found in free 
translational motion; it is always spontaneously pinned or frozen within
the ferromagnetic medium. (b) An FM bubble tends to move in a direction
perpendicular to an applied magnetic field gradient. For example, a single
bubble will undergo a $90^{\circ}$ deflection with respect to an externally
supplied uniform gradient in the absence of dissipation [5,$\!$ 6]. In the 
case of two or more interacting bubbles, a gradient arises intrinsically
and leads to a characteristic relative motion similar to the Hall motion
of electric charges in a uniform magnetic field or the motion of vortices
in a fluid [13].

Property (a) is clearly not the case for AFM bubbles or vortices because
of the Lorentz invariance of the underlying nonlinear $\sigma$ model.
Indeed, for any static vortex $\nb n=\nb n(\nb x)$ constructed in the 
preceding section, a vortex propagating freely with an arbitrary speed
$v<1(=c)$ in, say, the $x$-direction may be obtained by the elementary
Lorentz transformation
$$\nb n(x, y)\to\nb n\left({x-v\tau\over\sqrt{1-v^2}}, y\right) .\eqno (4.1)$$
One may then calculate the corresponding auxiliary fields from Eq. (2.19)
or (2.28), taking into account that the dynamical contribution
$(\nb n\times\dot{\nb n})$ is no longer vanishing, and subsequently
construct the spin configuration on the discrete lattice to obtain a rigidly
moving AFM vortex.

Similarly, there is no reason to believe that property (b) is sustained
for AFM bubbles or vortices. Yet one should expect that topology will
continue to play an important role within the relativistic dynamics.
A possible manifestation of a topological effect occurs in the dynamics of 
two bubbles in a head-on collision. Such a process was studied extensively
in the context of the isotropic O(3) nonlinear $\sigma$ model and shown
to exhibit a characteristic $90^{\circ}$ scattering pattern [14]. This
pattern is certainly unusual from the point of view of ordinary particle
dynamics but strongly reminiscent of the $90^{\circ}$ deflection of FM
bubbles in a field gradient. Of course, the above analogy is superficial but
indicates that both the Landau-Lifshitz and the relativistic dynamics are
influenced by the underlying topology.

The isotropic nonlinear $\sigma$ model is special in several ways.
In particular, its scale invariance leads to metastable bubbles
of arbitrary radius. The radius of each bubble changes during collision and
never returns to its initial value [14]. A healthier situation arises in 
the presence of anisotropy which sets a definite scale for the soliton size.
Then individual solitons may be deformed during scattering but will always
bounce back to their original shape well after collision. The AFM vortices
constructed in Sec. III are examples of 2D solitons with definite scale
and their dynamics will be the subject of the remainder of this paper.

The simplest dynamical experiment is to consider the evolution of a pair
of two idendical vortices which are initially at rest at a relative
distance $d$. Because vortices are extended structures, the above initial 
configuration is not uniquely defined and generally depends on the details
of the physical process that brings the two vortices to their initial
positions. However such details may be important for a short transient period 
but are not expected to significantly influence the global properties of 
the ensuing motion, especially when the initial distance $d$ is large. 
Therefore there exists significant freedom in the construction of the initial
configuration. The simplest choice is to consider the product ansatz
$$\Omega (x, y)=\Omega_{1}\left( x-{d\over 2}, y\right)
\Omega_{1}\left( x+{d\over 2}, y\right) ,\eqno (4.2)$$
where $\Omega$ is the complex stereographic variable
$$\Omega ={n_1+in_2\over 1+n_3}=\tan (\Theta/2)\, e^{i\Phi}\eqno (4.3)$$
for the vortex pair and $\Omega_{1}$ is the corresponding variable
for a single vortex. In view of Eq. (3.13) we may write
$$\Omega_{1}\left( x\pm {d\over 2}, y\right) =
{\sin\theta_{\pm}\over 1+\nu\cos\theta_{\pm}}e^{i\kappa (\phi_{\pm}+\phi_0)},
\eqno (4.4)$$
where
$$\theta_{\pm}=\theta (\rho_{\pm}),\qquad
\rho_{\pm}=\sqrt{\displaystyle{\left( x\pm {d\over 2}\right)^2+y^2}},\qquad
\phi_{\pm}=\hbox{arctan}{y\over x\pm{d\over 2}},\eqno (4.5)$$
and $\theta=\theta (\rho)$ is the profile of Fig. 3. In our simulations we
made the specific choices $\kappa =1$, $\nu =-1$ and $\phi_0=0$.
The product ansatz (4.2) is then written as
$$\Omega =f(\rho_+)e^{i\phi_+}f(\rho_-)e^{i\phi_-},\qquad
f(\rho_{\pm})=\hbox{cot}{\theta_{\pm}\over 2}\eqno (4.6)$$
and represents a pair of two identical vortices initially at rest,
at a relative distance $d$ on the $x$-axis.

The remaining steps for a complete specification of the initial
configuration on the discrete lattice of, say, Fig. 2 proceed as 
follows. The field $\nb n$ is obtained by inverting Eq. (4.3), i.e.,
$$n_1={\Omega +\overline{\Omega}\over 1+\overline{\Omega}\Omega},\qquad
n_2={1\over i}{\Omega -\overline{\Omega}\over 1+\overline{\Omega}\Omega},\qquad
n_3={1-\overline{\Omega}\Omega\over 1+\overline{\Omega}\Omega}.\eqno (4.7)$$

\noindent
The auxiliary fields $\nb m, \nb k$ and $\nb l$ are then computed from
Eq.$\!$ (2.28) applied for a static configuration 
$(\nb n\times\dot{\nb n}=0)$. The original spin variables 
$\nb A, \nb B, \nb C$ and $\nb D$ on a generic tetramer with coordinates
$(\alpha , \beta)$ are determined from Eq. (2.25) applied for the 
discrete set of points $x=2\varepsilon (\alpha -\alpha_0)$ and 
$y=2\varepsilon (\beta -\beta_0)$ of Eq.$\!$ (2.24) with 
$\alpha , \beta =1, 2, \ldots , N$. The spin configuration is thus specified
at every site of the original lattice.

Having determined the initial configuration, the ensuing evolution of the
vortex pair was calculated by a numerical solution of the initial-value
problem (2.2)-(2.4) using a fourth-order Runge-Kutta algorithm. Typical
runs were performed on a $200\times 200$ lattice for a reasonably weak
anisotropy $\varepsilon =0.1$ that leads to a respectable grid 
$[-10, 10]$ for the dimensionless position variables $x$ and $y$.
The rigidity of our numerical results was frequently checked on larger
lattices, up to $400\times 400$, and stability was improved by reenforcing
the constraint $\nb S_{i, j}^2=s^2$ at every site of the lattice after
every Runge-Kutta step. This numerical trick was borrowed from related
simulations in the isotropic nonlinear $\sigma$ model [14]. On this
occasion, it should be emphasized that we do not directly simulate the
dynamics of the $\sigma$ model but rather of the original Heisenberg
antiferromagnet. A byproduct of this fact was that various theoretical
predictions based on the continuum approximation of Sec. II were
verified in  detail. In particular, we have been able to illuminate
numerous subtle points in connection with the parity-breaking 
contributions in the auxiliary fields.

The results of the numerical simulation described above revealed no surprises
when the two vortices are initially at rest; they begin to drift away from
each other along the $x$-axis apparently in order to minimize their 
interaction energy. Therefore the calculated behavior is similar to that
of two ordinary particles interacting with a repulsive potential. 
Nevertheless this behavior is already significantly different from the one
observed in the case of two interacting FM bubbles which would rotate
around each other irrespectively of whether the potential is repulsive
or attractive [13].

A more interesting situation arises when the two vortices in Eq. (4.6) are
initially Lorentz boosted to velocities $\nb v$ and $-\nb v$ and are
thus set on a head-on collision course on the $x$-axis. As expected, the
two vortices begin to discelerate thanks to their mutual repulsion. 
However the future of the process depends crucially on the magnitude of 
the initial velocity. At low velocities the two vortices approach each
other to a minimum distance at which they come to rest and then turn around
and move off in opposite directions. When the initial speed exceeds a 
certain critical value the vortices again discelerate but come 
sufficiently close to a relative distance where the interaction potential 
seems to have become attractive. The vortices then begin to accelerate
toward each other until they overlap almost completely. More imporantly,
they subsequently split and reemerge as two separate vortices moving off
along the positive and negative $y$-direction, thus undergoing a 
$90^{\circ}$ scattering analogous to the one observed in numerical 
simulations of bubbles in the pure O(3) model [14]. The main difference
here is that vortices regain their initial shape except for a flip
of their internal phase equal to $\phi_0=\pm{\pi\over 2}$.

The situation described above is illustrated in Fig.$\!$ 5 which depicts three
characteristic snapshots of the collision of two like vortices that start
from an initial relative distance $d=5$ with velocities
$\nb v_1=(0.65, 0)$ and $\nb v_2=(-0.65, 0)$. We have chosen to illustrate
only the field $\nb n$, through its projection on the (12) plane, and 
hence the antiferromagnetic discontinuity present in Fig. 4 is not 
apparent in Fig. 5. However we have actually calculated the spin values
on the original lattice and then proceeded with the determination of 
$\nb n$ as well as the auxiliary fields. We were thus able to confirm
several important details of the formulation presented in Sec. II.

The picture changes drastically in the case of a vortex-antivortex
pair initially at rest, at a relative distance $d$ on the $x$-axis.
Such a pair is initially described by the product ansatz
$$\Omega =f(\rho_+)e^{i\phi_+}f(\rho_-)e^{-i\phi_-},\eqno (4.8)$$
where notation is the same as in Eq. (4.6). The subsequent evolution was
again studied by a numerical calculation similar to the one described
earlier for a pair of two like vortices. The vortex and the antivortex now
begin to accelerate toward each other apparently because their interaction
potential is attractive. Fig.$\!$ 6 depicts three characteristic snapshots
of the field $\nb n$ which indicate that the resulting spin configuration
shows no sign of a nontrivial topological structure, or that the 
vortex-antivortex pair is annihilated. Nevertheless a closer examination
of the time evolution of the energy density shown in Fig. 7 suggests
that the annihilation process is not completely dull. Indeed two distinct 
lumps of energy are emitted along the positive and negative $y$-axis
which do not correspond to any well defined soliton structures and eventually
dissipate into spinwaves. It is interesting that a tendency for a
$90^{\circ}$ scattering persists even in the present case [14].

To summarize, our calculation of head-on collisions of AFM vortices
confirms what appears to be a robust feature of relativistic topological
solitons [4,$\!$ 14]. It is more than
 clear that a simple theoretical explanation
of the observed behavior should be possible to obtain almost 
independently of the specific dynamical model. However such an explanation
has thus far been offered only within the context of some variant or other
of collective coordinates. For example, vortices in a Higgs model were
discussed along those lines in Ref.$\!$ [15] and shown to exhibit
$90^{\circ}$ scattering, at least in the so-called Bogomonly limit.
At this point, we do not have a convincing theoretical 
interpretation of our numerical data for head-on collisions. But the 
theoretical situation improves significantly in the case of AFM vortices
in a uniform magnetic field studied in the following section.

\vfill
\eject
\centerline{\ini V.\ VORTICES IN A MAGNETIC FIELD}
\bigskip
The effect of an applied uniform magnetic field $\nb H$ is accounted
for by including a Zeeman term in the Hamiltonian (2.1),
$$W\to W-g_0\mu_0\sum_{ij}(\nb H\cdot\nb S_{i, j}),\eqno (5.1)$$
where $g_0\sim 2$ is the gyromagnetic ratio and $\mu_0=e/2mc$ is the 
Bohr magneton divided by the Planck constant. AFM vortices then 
acquire the general dynamical features of FM bubbles described in the
beginning of Sec. IV. Such a radical change of behavior was already
speculated in Ref. [16] and is established here by means of 
unambiguous conservation laws that link the dynamics with the underlying
topology. The derived qualitative picture is then confirmed by direct
numerical simulations.

The classical ground state is now obtained by assigning a spin value
$\nb A$ on the first sublattice (solid circles) and a value $\nb B$
on the second (open circles). We further introduce the unit vectors
$\nb a=\nb A/s$ and $\nb b=\nb B/s$ to write for the energy of such
a configuration
$$W/s^2J\Lambda =2(\nb a\cdot\nb b)+{1\over 4}\varepsilon^2
(a^2_3+b^2_3)-{g_0\mu_0\over 2sJ}\nb H\cdot (\nb a+\nb b),\eqno (5.2)$$
where $\varepsilon$ is the dimensionless anisotropy constant of Eq. (2.5)
and $\Lambda$ is the total number of lattice sites assumed to be large.
For a field of strength $H$ applied along the third direction the minimum
of (5.2) is achieved by the canted spin configuration of Fig. 8, defined
up to an arbitrary azimuthal rotation, where the canting angle is given by
$$\sin\delta ={H\over H_c},\qquad H_c={(8+\varepsilon^2)sJ\over g_0\mu_0},
\eqno (5.3)$$
for field values in the range $H<H_c$. Above the critical field $H_c$
a transition takes place into a ferromagnetic phase 
$(\delta ={\pi\over 2})$. Actually we shall mostly study the parameter
regime where both the anisotropy and the applied field are weak, namely
$$\varepsilon\ll 1,\qquad H\ll H_c\approx {8sJ\over g_0\mu_0},\eqno (5.4)$$

\noindent
which are conditions for the validity of a continuum approximation and are
sufficiently nonstringent for practical applications. For future reference, 
Fig. 8 also displays the two vectors
$$\nb m={1\over 2}(\nb a+\nb b),\qquad \nb n={1\over 2}(\nb a-\nb b),
\eqno (5.5)$$
which may be expressed in terms of the canting angle as
$$\eqalign{\nb m & = (0, 0, \sin \delta)\approx
(g_0\mu_0 H/8sJ)(0, 0, 1),\cr
\noalign{\medskip}
\nb n & = (\cos \delta , 0, 0)\approx (1, 0, 0),\cr}\eqno (5.6)$$
where the second steps have been restricted to the parameter regime (5.4).
Therefore a nonvanishing magnetization $\nb m$ develops along the third
direction whereas $\nb n$ is confined in the basal plane.

When the field is turned on, the N\'{e}el state is set in a 
precessional mode that eventually relaxes into the canted state of Fig. 8
thanks to some dissipative process or other that is always present in a 
realistic antiferromagnet. Throughout this section we shall assume that
the field has been turned on sufficiently long to ensure that 
equilibrium has been achieved in the ground state. The argument is 
carried out in five steps described in the following five subsections.

\bigskip
\bigskip
\centerline{\bf A. The continuum model}
\bigskip
We now return to the discrete equations (2.6)-(2.7) which we extend
according to
$$\nb F_{\alpha , \beta}\to\nb F_{\alpha , \beta}+g_0\mu_0\nb H,\qquad
\nb G_{\alpha , \beta}\to\nb G_{\alpha , \beta}+g_0\mu_0\nb H.\eqno (5.7)$$
It is also convenient to introduce the rationalized field [3]
$$\nb h=h\nb e,\qquad h={g_0\mu_0 H\over 2\sqrt{2}\varepsilon sJ}.\eqno (5.8)$$
Eqs. (2.16) are then extended simply by adding a term
$\delta (\nb m\times\nb h)$ to the right-hand side of the first equation
and $\delta (\nb n\times\nb h)$ to the second, where
$\delta =\sqrt{2}\varepsilon$. The remaining algebraic details will be
omitted here because they are similar to those of Sec. II.

Hence the continuum limit is now governed by the strong inequalities
$$\varepsilon , \varepsilon h\ll 1,\eqno (5.9)$$

\noindent
the fields $\nb m$ and $\nb n$ continue to satisfy the constraints
$$\nb m\cdot\nb n=0,\qquad \nb n^2=1,\eqno (5.10)$$
the auxiliary field $\nb m$ is given by
$$\nb m={\varepsilon\over 2\sqrt{2}}[-(\nb n_{\eta}+\nb n_{\xi})+
(\nb n\times\dot{\nb n})-\nb n\times (\nb n\times\nb h)]\eqno (5.11)$$
and the field $\nb n$ satisfies the differential equation
$$\nb n\times\nb f=0,\qquad
\nb f=\ddot{\nb n}-\Delta\nb n+2(\nb h\times\dot{\nb n})+
(\nb n\cdot\nb h)\nb h+n_3\nb e.\eqno (5.12)$$
As a check of consistency one may explicitly verify that the ground state
values (5.6) are compatible with the above equations for parameters that 
satisfy the strong inequalities (5.9). Finally we note that Eq. (5.12)
emerges also in connection with the tetramerization scheme of Fig. 2,
while the associated auxiliary fields are
$$\eqalign{\nb m & = {\varepsilon\over 2\sqrt{2}}
[(\nb n\times\dot{\nb n})-\nb n\times (\nb n\times\nb h)],\cr
\noalign{\medskip}
\nb k & = -{\varepsilon\over 2}\nb n_x,\qquad
\nb l= -{\varepsilon\over 2}\nb n_y,\cr}\eqno (5.13)$$
which should be compared to Eqs. (2.28).

We may thus concentrate on the dynamics of the extended nonlinear $\sigma$
model (5.12) where the effect of the applied field is twofold; it breaks
Lorentz invariance and renormalizes the single ion anisotropy. An efficient
study of the dynamics is carried out through a variational principle, namely
$$\nb f=-{\delta{\cal A}\over \delta\nb n},\eqno (5.14)$$
where ${\cal A}$ is the action
$${\cal A}=\int L\, dxdyd\tau ,\eqno (5.15)$$
and $L$ is the corresponding Lagrangian density [2]
$$L={1\over 2}
[\dot{\nb n}^2-(\partial_{\mu}\nb n\cdot\partial_{\mu}\nb n)]+
\nb h\cdot (\nb n\times\dot{\nb n})-{1\over 2}
[(\nb n\cdot\nb h)^2+n^2_3].\eqno (5.16)$$
We further resolve the constraint $\nb n^2=1$ explicitly using, for example,
the spherical parametrization (3.3) to write

\noindent
$$\eqalign{L= & {1\over 2}[\dot{\Theta}^2+\sin^2\Theta\,\dot{\Phi}^2]-{1\over 2}
[(\partial_{\mu}\Theta\partial_{\mu}\Theta)+\sin^2\Theta\, 
(\partial_{\mu}\Phi\partial_{\mu}\Phi)]\cr
\noalign{\medskip}
& +  h\sin^2\Theta\,\dot{\Phi}-{1\over 2}(1+h^2)\cos^2\Theta.\cr}\eqno (5.17)$$
Hence there exist two pairs of canonical fields given by
$$\eqalign{\psi_1 = \Theta,\qquad \pi_1 & =\dot{\Theta}\cr
\noalign{\medskip}
\psi_2 = \Phi,\qquad \pi_2 & = \sin^2\Theta\, (\dot{\Phi}+h).\cr}\eqno (5.18)$$
The Hamiltonian is then obtained from
$$W=\int w\,  dx dy,\qquad w=\pi_a\dot{\psi}_a-L,\eqno (5.19)$$
where the repeated Latin index $a$ is summed over the two distinct
values of Eq.$\!$ (5.18). A more explicit form of the energy density $w$
expressed directly in terms of the field $\nb n$ reads
$$w={1\over 2}[\dot{\nb n}^2+(\partial_{\mu}\nb n\cdot\partial_{\mu}\nb n)]
+{1\over 2}(1+h^2)n^2_3,\eqno (5.20)$$
where we find no trace of the nonrelativistic term 
$\nb h\cdot (\nb n\times\dot{\nb n})$ of Eq. (5.16). The energy $W$ is now
measured in units of $2\sqrt{2}\varepsilon s^2J$.

Therefore the extended nonlinear $\sigma$ model (5.12) may be cast in the
standard Hamiltonian form
$$\dot{\psi}_a={\delta W\over\delta\pi_a},\qquad
\dot{\pi}_a=-{\delta W\over\delta\psi_a};\qquad
a=1, 2, \eqno (5.21)$$
which will be the basis for our subsequent theoretical discussion. For
instance, the conserved linear momentum should be given by
$$p_{\mu}=-\int \pi_a\partial_{\mu}\psi_a\, dx dy,\eqno (5.22)$$
and the angular momentum by 
$$\ell =-\int\pi_a\varepsilon_{\mu\nu}x_{\mu}\partial_{\nu}\psi_a\, 
dx dy.\eqno (5.23)$$
Using the canonical variables of Eq. (5.18) a more explicit form of linear
momentum reads
$$p_{\mu}=-\int (\dot{\Theta}\partial_{\mu}\Theta +\sin^2\Theta\, 
\dot{\Phi}\partial_{\mu}\Phi +h \sin^2\Theta\,  \partial_{\mu} \Phi)
dx dy\eqno (5.24)$$
and is in agreement with the expression quoted in Ref.$\!$ [16]. However
the field dependent term in Eq.$\!$ (5.24) leads to an improper integral
for vortex configurations for which $\Theta\sim{\pi\over 2}$ and 
$\Phi\sim\kappa\phi$ at spatial infinity. One would think that this 
difficulty may be resolved by modifying the momentum density by a total
divergence, thus replacing $\sin^2\Theta\, \partial_{\mu}\Phi$ by
$(\sin^2\Theta -1)\partial_{\mu}\Phi=-\cos^2\Theta\, \partial_{\mu}\Phi$,
which would indeed lead to proper behavior at infinity where 
$\cos\Theta =0$. Nevertheless the ambiguity would then be shifted to the
origin of the vortex where $\partial_{\mu}\Phi$ is singular, because
$\Phi$ is multivalued, while $\Theta =0$ or $\pi$ and 
$\cos\Theta =\pm 1\not =0$.

The ambiguities in the linear momentum signal an important link between
the dynamics and the underlying topological complexity, in analogy with 
the situation previously analyzed for FM bubbles [5,$\!$ 6]. In turn,
the relativistic dynamics of AFM vortices studied in earlier sections
should be radically altered by the applied field. A complete resolution
of the ambiguities is given in subsection C after the ground is prepared
in subsection B.

\bigskip
\bigskip
\centerline{\bf B. Vorticity and the stress tensor}
\bigskip
For any 2D field theory that can be brought to the standard 
Hamiltonian form (5.22) one may define a (scalar) vorticity
$$\gamma =\varepsilon_{\mu\nu}\partial_{\mu}\pi_a\partial_{\nu}\psi_a,
\eqno (5.25)$$
where $\varepsilon_{\mu\nu}$ is the 2D antisymmetric tensor 
$(\varepsilon_{11}=0=\varepsilon_{22}$, 
$\varepsilon_{12}=1=-\varepsilon_{21})$. Terminology is borrowed from fluid
dynamics because the quantity $\gamma$ shares with ordinary vorticity
several formal properties. The time derivative of $\gamma$ is calculated
from the Hamilton equations to yield
$$\dot{\gamma}=-\varepsilon_{\mu\nu}\partial_{\mu}\tau_{\nu}\eqno (5.26)$$
where the vector density
$$\tau_{\nu}={\delta W\over\delta\psi_a}\partial_{\nu}\psi_a+
{\delta W\over\delta\pi_a}\partial_{\nu}\pi_a\eqno (5.27)$$
is analogous to the ``force density'' employed by Thiele [17] in the problem
of FM bubbles. It is not difficult to see that $\tau_{\nu}$ may be
written as a total divergence,
$$\tau_{\nu}=\partial_{\lambda}\sigma_{\nu\lambda},\eqno (5.28)$$
where the tensor $\sigma_{\nu\lambda}$ is the stress tensor
$$\sigma_{\nu\lambda}=w\delta_{\nu\lambda}-
{\partial w\over\partial (\partial_{\lambda}\psi_a)}\partial_{\nu}\psi_a-
{\partial w\over\partial (\partial_{\lambda}\pi_a)}\partial_{\nu}\pi_a
\eqno (5.29)$$
calculated for a specific energy density $w$. Eq. (5.26) then reads
$$\dot{\gamma}=-\varepsilon_{\mu\nu}\partial_{\mu}
\partial_{\lambda}\sigma_{\nu\lambda}\eqno (5.30)$$
and proves to be fundamental for our purposes [5,6].

It should be noted that the preceding discussion makes no distinction
between ordinary field theories and those endowed with nontrivial 
topological structure or related properties. However a clear
distinction emerges when we consider the total vorticity
$$\Gamma =\int\gamma\,  dxdy=\varepsilon_{\mu\nu}\int
\partial_{\mu}\pi_a\partial_{\nu}\psi_a\,  dx dy,\eqno (5.31)$$
which is conserved by virtue of Eq. (5.30) for any field configuration 
with reasonable behavior at infinity. One may also write
$$\Gamma =\varepsilon_{\mu\nu}\int
[\partial_{\mu}(\pi_a\partial_{\nu}\psi_a)-\pi_a
\partial_{\mu}\partial_{\nu}\psi_a]dxdy\eqno (5.32)$$
to indicate that a vanishing value of the total vorticity is the rule rather
than the exception. Indeed, under normal circumstances, the first term in 
(5.32) is shown to vanish by transforming it into a surface integral at
infinity and the second term also vanishes because 
$\varepsilon_{\mu\nu}\partial_{\mu}\partial_{\nu}\psi_a=0$ for any
differentiable function $\psi_a$. Yet the above conditions may not be met
in a field theory with nontrivial topology, a fact closely related to the
ambiguities discussed in connection with the linear momentum. In general,
the canonical definition of conservation laws is rendered ambiguous
when the total vorticity $\Gamma$ is different from zero.

It is then important to examine more closely the definition of vorticity
in the current model. Substitution of the canonically conjugate fields of
Eq.$\!$ (5.18) in Eq.$\!$ (5.25) and straightforward algebraic manipulation
yield the local vorticity
$$\gamma =\varepsilon_{\mu\nu}\partial_{\mu}
(\dot{\nb n}\cdot\partial_{\nu}\nb n)+h\omega ,\eqno (5.33)$$
where
$$\omega =-{1\over 2}[\varepsilon_{\mu\nu}\sin (2\Theta)\, \partial_{\nu}
(2\Theta)\partial_{\mu}\Phi].\eqno (5.34)$$
The first term in Eq. (5.33) is an uncomplicated total divergence which
leads to a vanishing contribution in the total vorticity $\Gamma$
of Eq. (5.31). Thus we may write
$$\Gamma =h\int \omega\, dx dy\eqno (5.35)$$
and further note that the density $\omega$ of Eq. (5.34) resembles the
Pontryagin density $q$ of Eq. (3.18) except for an overall factor
$-{1\over 2}$ and the replacement $\Theta\to 2\Theta$. The latter
suggests considering the three-component vector
$\nb N=(N_1, N_2, N_3)$ with
$$\eqalign{N_1 & = 2n_3n_1=\sin (2\Theta)\cos\Phi ,\cr
\noalign{\medskip}
N_2 & = 2n_3n_2=\sin (2\Theta)\sin\Phi ,\cr
\noalign{\medskip}
N_3 & = 2n_3-1=\cos (2\Theta),\cr}\eqno (5.36)$$
which is also a unit vector field $(\nb N^2=1)$. The density $\omega$
may then be written as
$$\omega =-{1\over 4}\varepsilon_{\mu\nu}
(\partial_{\nu}\nb N\times\partial_{\mu}\nb N)\cdot\nb N\eqno (5.37)$$
and should be compared to the standard Pontryagin density of Eq. (3.17).
Furthermore the field $\nb N$ satisfies the simple boundary condition
$$\nb N\mathop{\longrightarrow}\limits_{|\nb x|\to\infty}
(0, 0, -1),\eqno (5.38)$$
thanks to the condition $n_3\to 0$ satisfied by all relevant field
configurations, including the vortex configurations of Sec.$\!$ III.
The net conclusion is that $\omega$ is actually the Pontryagin density
for the field $\nb N$ and thus yields an integer-valued total vorticity
$$\Gamma =2\pi h\kappa ,\qquad \kappa =0, \pm 1, \pm 2, \ldots , \eqno (5.39)$$
where the integer $\kappa$ will be referred to as the vortex number.
Indeed an explicit calculation for a single vortex or antivortex
$(\kappa =\pm 1)$ discussed in Sec.$\!$ III confirms Eq.$\!$ (5.39)
 for any choice
of the polarity $\nu$, in contrast to Eq. (3.20) that depends on both the 
vortex number and the polarity.

To complete this level of description of the current model we quote an 
explicit expression for the stress tensor calculated from Eq. (5.29) using
as input the energy density (5.20). The final result is
$$\sigma_{\nu\lambda}=w\delta_{\nu\lambda}-
(\partial_{\nu}\nb n\cdot\partial_{\lambda}\nb n),\eqno (5.40)$$
where the energy density $w$ may be expressed directly in terms of the
field $\nb n$ through Eq. (5.20).

Finally we mention that the discussion of this subsection possesses a 
straightforward 3D generalization. For instance, Eq. (5.30) becomes
$$\dot{\gamma}_i=-\varepsilon_{ijk}\partial_j\partial_l
\sigma_{kl},\eqno (5.41)$$
where Latin indices $i, j, \ldots$ assume three distrinct values and 
$\varepsilon_{ijk}$ is the 3D antisymmetric tensor. The stress tensor
in Eq. (5.41) is obtained by an obvious 3D extension of Eq. (5.40) and
the vorticity $\hbox{\bf ã}=(\gamma_1, \gamma_2, \gamma_3)$ is given by
$$\gamma_i=\varepsilon_{ijk}\partial_j\pi_a\partial_k\psi_a=
\varepsilon_{ijk}\partial_j(\dot{\nb n}\cdot\partial_k\nb n)+
h\omega_i,\eqno (5.42)$$

\noindent
where the vector density $\hbox{\bf ù}=(\omega_1, \omega_2, \omega_3)$
reads 
$$\omega_i=-{1\over 4}\varepsilon_{ijk}(\partial_k\nb N\times
\partial_j\nb N)\cdot\nb N\eqno (5.43)$$
and generalizes the scalar Pontryagin density (5.37). Furthermore the
vorticity is solenoidal $(\bnabla\cdot\hbox{\bf ã}=0)$ essentially by
construction and may thus be derived from a vector potential 
$(\hbox{\bf ã}=\bnabla\times\nb a)$ which can be used to define a degree
of knottedness or helicity of vortex lines,
$${\cal H}\sim \int (\nb a\cdot\hbox{\bf ã})dV,\eqno (5.44)$$
by analogy with the Hopf index in ferromagnets [6]. Such an index may prove
to be an important issue in the study of 3D antiferromagnets but will
not be discussed further in the present paper.

\bigskip
\bigskip
\centerline{\bf C. Conservation laws}
\bigskip
We now return to the 2D theory and consider the derivation of unambiguous
conservation laws. Since the main strategy was already explained in the
related context of FM bubbles [5,6] our description here will address only
the essential points adapted to the present model. The appearance of a 
double derivative in the right-hand side of the fundamental relation (5.30)
suggests that some of the low moments of the local vorticity $\gamma$
must be conserved. Indeed the linear momentum $\nb p=(p_1, p_2)$ is given by
$$p_{\mu}=-\varepsilon_{\mu\nu}I_{\nu},\qquad
I_{\nu}=\int x_{\nu}\gamma\,  dx dy,\eqno (5.45)$$
and the angular momentum $\ell$ by
$$\ell ={1\over 2}\int\rho^2\gamma\,  dx dy,\eqno (5.46)$$
where $\rho^2=x^2+y^2$. The list of conservation laws is completed by the
total magnetization $\mu$ in the third direction,
$$\mu =\int [\nb e\cdot (\nb n\times\dot{\nb n})-hn^2_3] dx dy,\eqno (5.47)$$
which can be derived directly from the equation of motion.

The preceding identifications are made plausible by inserting the general
expression for the vorticity given by Eq.$\!$ (5.25) in Eqs.$\!$ (5.45)
 and (5.46)
and by freely performing partial integrations to recover the canonical
forms of linear and angular momentum quoted in Eqs. (5.22) and (5.23)
 which are
plagued by the ambiguities discussed in connection with Eq. (5.24). However
no such ambiguities occur in Eqs.$\!$ (5.45) and (5.46) because the local 
vorticity $\gamma$ can be obtained directly from the field $\nb n$,
rather than the angular variables, and is a particularly well defined
quantity; see Eqs.$\!$ (5.35) and (5.37). In other words, partial integrations
should be performed with great care and are often unjustified.

The main point of this theoretical exercise is that the very structure of the
conservation laws (5.45)-(5.47) suggests a radical change in the dynamical
behavior of vortices in an applied field $(h\not =0)$. The effect of a
nonvanishing total vorticity $(\Gamma =2\pi h\kappa)$ becomes apparent
by considering the transformation of the moments $I_{\nu}$ of Eq. (5.45)
under a translation of coordinates 
$\nb x\to\nb x+\nb c$ where $\nb c=(c_1, c_2)$ is a constant vector,
$$I_{\nu}\to I_{\nu}+\Gamma c_{\nu},\eqno (5.48)$$
which implies a nontrivial transformation of the linear momentum (5.45)
when $\Gamma\not =0$. This is surely an unusual property, because linear
momentum should be expected to remain unchanged under a constant 
translation of the origin of coordinates, and indicates that the moments
$I_{\nu}$ provide a measure of position rather than momentum. Such a fact
is made explicit by considering the guiding center vector
$\nb R=(R_1, R_2)$ with coordinates
$$R_{\nu}={I_{\nu}\over\Gamma}={1\over\Gamma}\int x_{\nu}
\gamma\,  dx dy,\eqno (5.49)$$
which transforms as $\nb R\to\nb R+\nb c$ under a constant translation and
is thus a measure of position of a spin configuration with $\Gamma\not =0$.
Nevertheless the vector $\nb R$ is conserved.

A related fact is that the familiar Poisson bracket algebra is significantly
affected when $\Gamma\not =0$. Using the canonical Poisson brackets,
$$\{\pi_a(\nb x),\, \psi_b(\nb x')\} =\delta_{ab}
\delta (\nb x-\nb x'),\eqno (5.50)$$
and the general expression of the local vorticity (5.25) in the definition
of the linear momentum (5.45), it is not difficult to establish the relations
$$\{ p_1, p_2\} =\Gamma,\qquad \{ R_1, R_2\} =1/\Gamma,\eqno (5.51)$$
which are strongly reminiscent of the situation in the case of electron
motion in a uniform magnetic field, the role of the latter being played
here by the total vorticity $\Gamma$.

Similarly the angular momentum (5.46) actually provides a measure of the
vortex size, a fact made explicit by considering the mean squared radius
defined from
$$r^2={1\over\Gamma}\int (\nb x-\nb R)^2\gamma\, dx dy={2\ell\over\Gamma}
-\nb R^2,\eqno (5.52)$$
which is also conserved. Needless to say, the conservation laws (5.45) and 
(5.46) resume their ordinary physical significance at vanishing total
vorticity $(\Gamma =0)$.

The observed transmutation in the physical significance of the conservation
laws of linear and angular momentum implies a radical change in the
dynamical behavior of topological solitons. For example, a single AFM
vortex or antivortex $(\kappa =\pm 1)$ in a uniform magnetic field carries
a nonvanishing total vorticity $(\Gamma =\pm 2\pi h)$ and thus cannot
be found in a free translational motion $(\dot{\nb R}=0)$. It is always
spontaneously pinned or frozen within the antiferromagnetic medium, in 
contrast to the freely moving vortices occuring in the relativistic theory
at vanishing field. Vortex motion can occur in the presence of other vortices,
but the dynamical pattern is also expected to be substantially different
from the one obtained at vanishing field in Sec.$\!$ IV. Specifically, 
interacting AFM vortices should now behave as ordinary vortices in a fluid
or as electric charges in a uniform magnetic field, as demonstrated by
direct simulations in subsection D.

The preceding descussion was kept deliberately general in order to 
emphasize that the emerging qualitative picture is valid in any field theory
for which the total vorticity $\Gamma$ may be different from zero. However
it is now useful to express the conservation laws in a more explicit
form that takes into account the specific structure of the current model.
We thus insert the local vorticity of Eq. (5.33) in Eq. (5.45) to
obtain the linear momentum
$$p_{\mu}=-\int [(\dot{\nb n}\cdot\partial_{\mu}\nb n)+h
\varepsilon_{\mu\nu}x_{\nu}\omega] dx dy,\eqno (5.53)$$
where we have performed a partial integration in the first term which is
free of all ambiguities. Similarly the angular momentum (5.46) reads
$$\ell =\int [-\varepsilon_{\mu\nu}x_{\mu}
(\dot{\nb n}\cdot\partial_{\nu}\nb n)+{1\over 2}h\rho^2\omega]
dx dy.\eqno (5.54)$$
As mentioned already, the above conservation laws possess their usual
physical significance only at vanishing total vorticity,
$\Gamma =2\pi h\kappa =0$, which may be achieved when either the applied
field $h$ or the vortex number $\kappa$ vanishes. Otherwise one must
consider the guiding center coordinates (5.49) or
$$R_{\mu}={1\over 2\pi h\kappa}\int [-\varepsilon_{\mu\nu}
(\dot{\nb n}\cdot\partial_{\nu}\nb n)+h x_{\mu}\omega]dx dy\eqno (5.55)$$
and the radius $r$ calculated from Eq. (5.52). It should be noted that the
preceding conservation laws display some formal similarities to those 
derived in a model for a superconductor [18,19].

Finally we return briefly to the 3D theory discussed in the concluding
paragraph of subsection B and quote the corresponding conservation laws
of linear and angular momentum
$$\nb p=-{1\over 2}\int (\nb r\times\hbox{\bf ã})dV,\qquad \nb l
 =-{1\over 3}\int [\nb r\times (\nb r\times\hbox{\bf ã})]dV,\eqno (5.56)$$
where $\nb r=(x, y, z)$, $dV=dx dy dz$ and 
$\hbox{\bf ã}=(\gamma_1, \gamma_2, \gamma_3)$ is the vector vorticity
field of Eq. (5.42). It is interesting that (5.56) are formally idendical
to the conservation laws derived in fluid dynamics, at least for 
incompressible fluids; see Eqs. (7.2.5) and (7.2.6) of Ref. [20].

\bigskip
\bigskip
\centerline{\bf D. Interacting vortices}
\bigskip
In the presence of a bias field the static vortices of Sec. III adjust
to a slightly different shape. It is not difficult to see that the 
functional form of a static vortex remains the same as in Eq. (3.13) 
except that $\theta =\theta (\rho)$ now satisfies the ordinary 
differential equation
$${1\over\rho}{\partial\over\partial\rho}
\left(\rho{\partial\theta\over\partial\rho}\right) +
\left( 1+h^2-{1\over\rho^2}\right)\cos\theta\sin\theta =0,\eqno (5.57)$$
which differs from Eq. (3.10) only by an additional easy-plane
anisotropy with strength equal to $h^2$. Consequently Eq.$\!$ (5.57)
reduces to Eq.$\!$ (3.10) by the simple rescaling
$\overline{\rho}=\sqrt{1+h^2}\rho$ and the vortex profile
$\theta =\theta (\overline{\rho})$ is again given by Fig. 3 with the
replacement $\rho\to\overline{\rho}$. More importantly, the auxiliary
fields (5.11) or (5.13) now contain field dependent terms that are
crucial for a correct calculation of the actual spin values on the 
lattice of Fig.$\!$ 1 or Fig.$\!$ 2, respectively.

A pair of like vortices initially at rest is described by the product
ansatz (4.2) taking into account the field dependent modifications 
discussed in the preceding paragraph. The ensuing time evolution of the
vortex pair was obtained numerically. Instead of drifting away the two
vortices actually begin to rotate around each other, in sharp contrast to
the situation described in Sec. IV at vanishing field. In other words,
each vortex moves in a direction perpendicular to the applied force,
in analogy with the skew deflection of FM bubbles in a field gradient [5,6].
Fig.$\!$ 9 illustrates the
 initial configuration together with two characteristic
snapshots taken at time intervals such that the pair had rotated roughly 
by $90^{\circ}$ and $180^{\circ}$, respectively. It is interesting to note
that a $90^{\circ}$ rotation of the pair in real space is always followed
by a $90^{\circ}$ internal phase shift of each vortex. Fig.$\!$ 10 depicts
the actual trajectories obtained by tracking the points where
$|n_3|=1$. Inspite of an apparent initial tendency to drift away, the two
vortices eventually orbit around each other, in complete analogy with
the 2D motion of two like vortices in an ordinary fluid or two interacting
electrons in a uniform magnetic field. The observed departures of the 
trajectories of Fig. 10 from a circular shape correspond to the 
well-known Larmor oscillations in the electron problem. These
oscillations are expected to be smoothed out in the limit of large
relative distance.

The above results are consistent with the qualitative picture
suggested by the conservation laws of subsection C.
The two-vortex system carries a total vortex number $\kappa =2$ and thus
a nonvanishing total vorticity $\Gamma =4\pi h$. The guiding center
calculated from Eq.$\!$ (5.55) is initially located at the origin of the 
coordinate system and remains fixed at all later times. The angular
momentum was calculated numerically based on Eq. (5.54) and its time
evolution is demonstrated in Fig.$\!$ 11. Although the two pieces of
 Eq.$\!$ (5.54)
acquire a nontrivial time dependence, their sum is fairly well conserved.
Furthermore the same general picture was obtained by repeating the
calculation for a pair of vortices with the same vortex numbers
$(\kappa_1=\kappa_2)$ but opposite polarities $(\nu_1=-\nu_2)$. This
result is consistent with the fact that the driving issue is the total
vorticity $\Gamma =2\pi h(\kappa_1+\kappa_2)$ which is independent of
the polarities, in contrast to the ordinary winding number 
$Q$ of Eq. (3.20).

The calculation was further repeated for a vortex-antivortex pair
initially described by the product ansatz (4.8) incorporating the
appropriate field dependent modifications. Recall that at vanishing
field the vortex and the antivortex are attracted toward each other and
are eventually annihilated. The situation is drastically different at
nonvanishing field. The pair undergoes Kelvin motion roughly along
parallel lines that are perpendicular to the line connecting the
vortex and the antivortex. Fig.$\!$ 12 depicts the initial configuration
together with a snapshot taken at a later instance when the pair
had moved in formation along the $y$-axis to a distance approximately
equal to the initial relative separation. Fig. 13 demonstrates the actual
trajectories (solid lines) obtained by tracking the points where
$|n_3|=1$. Therefore the derived picture is qualitatively identical to 
the Kelvin motion of a vortex-antivortex pair in an ordinary fluid or
the Hall motion of an interacting electron-positron pair in a uniform
magnetic field.

Returning to the conservation laws we note that the total vorticity
of a vortex-antivortex pair vanishes 
$(\kappa =\kappa_1+\kappa_2=0$ and hence $\Gamma =0)$. Therefore it is now
meaningful to interpret (5.53) as the conserved total linear momentum
of the system. A nonvanishing component develops only along the 
$y$-axis, i.e., along the direction of motion of the pair, and its
conservation is demonstrated in Fig. 14. Again each of the two pieces
in Eq. (5.53) exhibits a nontrivial time dependence but their sum is
fairly well conserved. On the other hand, it is still meaningful
to define individual guiding centers when the pair is widely separated.
For example, approximate guiding centers for the vortex and the
antivortex may be defined by restricting the integration in Eq. (5.55) to
the right and left half plane and setting $\kappa =1$ and $\kappa =-1$,
respectively. The trajectories obtained by tracking the above
approximate guiding centers are also shown in Fig. 13 (dashed lines)
and are close to two parallel straight lines. The analogy with 
the motion of an electron-positron pair in a uniform magnetic field
in now made more definite. The actual trajectories of the electron and 
positron undergo Larmor oscillations along the parallel trajectories
of their guiding centers, which become increasingly narrower with 
increasing relative separation. The absence of more than one such
oscillation in Fig. 13 is due to our (numerical) inability to follow the
motion to a larger distance; see, however, a related calculation of 
interacting vortices in a charged fluid [19]. Finally we have verified
that a vortex-antivortex pair $(\kappa_1+\kappa_2=0)$ exhibits Kelvin
motion for any choice of relative polarities 
$(\nu_1=\nu_2$ or $\nu_1=-\nu_2)$, a fact that reenforces the prominence
of the total vorticity $\Gamma$ in the study of dynamics.

To summarize, a simple comparison of the results of this subsection
to those of Sec. IV establishes that the dynamics of AFM vortices
is profoundly altered by the applied field, in remarkable analogy
with the familiar Hall effect. Nevertheless one should keep in mind
that the effect of a bias field on AFM vortices would not have been
as drastic without the aid of the underlying nontrivial topological
structure.

\bigskip
\bigskip
\centerline{\bf E.\ The isotropic antiferromagnet}
\bigskip
The special case of an isotropic Heisenberg antiferromagnet is important
for both practical and theoretical purposes. The isotropic limit was 
briefly discussed in the concluding paragraphs of Sec.$\!$ III in the absecne
of a bias field. It was then mentioned that the model possesses metastable
AFM bubbles, instead of vortices, which are characterized by the standard
Pontryagin index (3.17). However, when a bias field is turned on, the
picture changes drastically for two reasons. First, the applied field
itself supplies an effective easy plane anisotropy that leads to vortices
instead of bubbles. Second, the dynamics of vortices departs 
significantly from the relativistic dynamics of the pure
antiferromagnet.

In the remainder of this section we shall briefly describe the necessary
modifications of the formalism to accomodate the isotropic model in a 
uniform magnetic field. Since a single ion anisotropy is no longer
available to provide the small parameter $\varepsilon$ of Eq.$\!$ (2.5),
such a parameter is now furnished by the applied field which is assumed
to be weak:
$$\varepsilon ={g_0\mu_0H\over 2\sqrt{2}sJ}\ll 1.\eqno (5.58)$$
The relevant dynamical equations are then obtained from our earlier results
by the formal substitution $\nb h\to \nb e=(0, 0, 1)$, or $h\to 1$,
and by omitting the contribution from the single ion anisotropy.
Thus the dynamics of the field $\nb n$ is now governed by the parameter
free Lagrangian
$$L={1\over 2}[\dot{\nb n}^2-(\partial_{\mu}\nb n\cdot\partial_{\mu}\nb n)]
+\nb e\cdot (\nb n\times\dot{\nb n})-{1\over 2}(\nb e\cdot\nb n)^2,
\eqno (5.59)$$
which leads to the equation of motion
$$\nb n\times\nb f=0,\qquad \nb f=\ddot{\nb n}-\Delta\nb n+2
(\nb e\times\dot{\nb n})+(\nb n\cdot\nb e)\nb e.\eqno (5.60)$$
The associated auxiliary fields are accordingly given by
$$\nb m={\varepsilon\over 2\sqrt{2}}[-(\nb n_{\eta}+\nb n_{\xi})+
(\nb n\times\dot{\nb n})-\nb n\times (\nb n\times\nb e)],\eqno (5.61)$$
for the lattice of Fig. 1, or
$$\eqalign{\nb m & = {\varepsilon\over 2\sqrt{2}}
[(\nb n\times\dot{\nb n})-\nb n\times (\nb n\times\nb e)],\cr
\noalign{\medskip}
\nb k & = -{\varepsilon\over 2}\nb n_x,\qquad
\nb l= -{\varepsilon\over 2}\nb n_y,\cr}\eqno (5.62)$$
for the lattice of Fig. 2. Finally we must set $h=1$ throughout our
discussion of conservation laws.

Therefore the corresponding physical picture can be readily inferred
without further calculation. Static vortices are formally identical
to those of Sec.$\!$ III but their dynamics is similar to the
nonrelativistic dynamics of the current Sec.$\!$ V, for any finite value
of the applied field. In other words, to the extent that topological 
solitons are relevant for the physics of an isotropic antiferromagnet,
the dynamical picture is changed significantly even by a very weak bias
field.

\bigskip
\bigskip
\centerline{\ini VI.\ CONCLUDING REMARKS}
\bigskip
The emphasis in the main text was placed on elementary processes involving
only two AFM vortices, in order to clearly illustrate an important link
between topology and dynamics. However further progress in that direction
hinges upon the actual production of isolated vortices. Although there exist
several examples of realistic antiferromagnets that are effectively
two-dimesnional, including the parent compounds of high-$T_c$
superconductors, there seems to be no direct experimental evidence for 
isolated AFM vortices or bubbles. This situation is in marked contrast
to the observed abundance of ferromagnetic bubbles, vortices in superfluid
helium, or Abrikosov vortices in superconductors, and may change in 
the future.

The qualitative picture derived from elementary processes must also 
influence the thermodynamics of 2D antiferromagnets. Work in that 
direction was already presented in Refs.$\!$ [21,$\!$ 22] for both Ising-like
and single-ion anisotropy, while the effect of an applied field was
considered in Ref.$\!$ [16]. It is clear that much remains to be done in 
connection with the anticipated Berezinskii-Kosterlitz-Thouless
(BKT) phase transition which relies on the dynamics of a gas of 
interacting vortices and antivortices. Suffice it to say that the
dynamics of vortex-antivortex pairs studied in Sec. IV is 
radically modified by an applied magnetic field discussed in Sec.$\!$ V.
The BKT theory may have to be reformulated in a way that clearly
reflects the fundamental change of behavior in the elementary vortex
processes when a field is turned on.

Perhaps the clearest manifestation of the effect of an applied field will
emerge in the thermodynamics of an isotropic antiferromagnet. Topological
solitons at vanishing field are metastable AFM bubbles that obey relativistic
dynamics. However the smallest external field will trigger Hall dynamics
for AFM vortices which become the relevant topological excitations. Again
a successful BKT theory must reflect this abrupt transition in the limit
of vanishing field.

Our discussion is concluded with some comments on a variation of the main
theme that has not been treated in this paper. An easy-axis anisotropy
$(g<0)$ would lead to a ground (N\'{e}el) state that is polarized along the
easy axis. Therefore topological solitons must then satisfy the simple
boundary condition (3.16) and would be AFM bubbles classified by the standard
Pontryagin index (3.17). However an application of the Derrick theorem either
in its original form or its extended version discussed in the Appendix
leads to the conclusion that such solitons do not exist either with finite
or infinite energy. This is a notable difference from FM bubbles that occur
in easy-axis ferromagnets. The latter are stabilized by a combination
of the effects of the long-range magnetostatic field created in a 
ferromagnetic film and of an applied bias field [6].

Nevertheless, when an easy-axis antiferromagnet is immersed in a uniform
magnetic field pointing along the symmetry axis, an effective easy-plane
anisotropy is produced that competes with the easy-axis anisotropy. And,
when the field exceeds a certain critical value, a spin-flop transition
takes place from the N\'{e}el state polarized in the third direction
to a canted state of the type shown in Fig. 8 which exhibits
azimuthal degeneracy. Consequently AFM vortices reappear above the critical
value of the applied field and their dynamical properties are very
similar to those discussed in Sec. V.

\bigskip
\bigskip
\centerline{\ini ACKNOWLEDGMENTS}
\bigskip
We are grateful to W.J. Zakrzewski for his hospitality at Durham and for
discussion of some of the issues presented in this paper. The work was
supported in part by a grant from the EEC(CHRX-CT93-0332) and by a
bilateral Greek-Slovak research program.

\bigskip
\bigskip
\centerline{\ini APPENDIX: VIRIAL THEOREMS}
\bigskip
We derive a slight generalization of the Derrick theorem [7] that is not
based on the assumption of finite soliton energy. The method is an 
elementary extension of our earlier work on magnetic bubbles [6].
We consider only static solutions for which the stress tensor (5.40)
in the absence of an external field reduces to
$$\eqalign{\sigma_{\nu\lambda} & = w\delta_{\nu\lambda}-
[(\partial_{\nu}\Theta\partial_{\lambda}\Theta)+\sin^2\Theta\, 
(\partial_{\nu}\Phi\partial_{\lambda}\Phi)],\cr
\noalign{\medskip}
w & = {1\over 2}[(\partial_{\mu}\Theta\partial_{\mu}\Theta)+\sin^2\Theta\, 
(\partial_{\mu}\Phi\partial_{\mu}\Phi)+\cos^2\Theta],\cr}\eqno 
\hbox{(A.1)}$$
and satisfies the continuity equation
$$\partial_{\lambda}\sigma_{\nu\lambda} =0,\eqno \hbox{(A.2)}$$
on account of the static Hamilton equations (3.5).

A series of virial relations may now be derived by taking suitable
moments of (A.2), the simplest possibility being
$$\int_Ax_{\nu}\partial_{\lambda}\sigma_{\mu\lambda}
dx dy =0,\eqno \hbox{(A.3)}$$
where the integration extends over some finite region A in the 2D plane.
An application of the divergence theorem yields
$$\int_A\sigma_{\mu\nu} dx dy=\oint_Cx_{\nu}
\sigma_{\mu\lambda} ds_{\lambda},\eqno \hbox{(A.4)}$$
where the integral in the right-hand side is taken over the closed
curve $C$ surrounding the area A; by convention the line-element vector
$(ds_1, ds_2)$ is perpendicular to the curve. On the assumption that
all fields approach asymptotic values such that the volume integrals in
the left-hand side are finite and the surface integrals in the right-hand
side vanish, when the integration extends to infinity, one obtains
the virial relations
$$\int\sigma_{\mu\nu} dx dy=0,\eqno \hbox{(A.5)}$$
where $\mu , \nu =1$ or 2 are taken in any combination. It is not 
difficult to see that relations (A.5) may also be obtained by a 
Derrick-like argument where coordinates are rescaled according to 
$x\to\lambda_{11}x+\lambda_{12}y$, $y\to\lambda_{21}x+\lambda_{22}y$
and we further demand that the resulting total energy be stationary
at $\lambda_{11}=1=\lambda_{22}$ and $\lambda_{12}=0=\lambda_{21}$.
In fact, the original scaling relation of Derrick may be recovered by taking
the trace of Eq. (A.5):
$$\int tr\sigma\, dx dy=0.\eqno \hbox{(A.6)}$$
Substitution in Eq.$\!$ (A.6) of the explicit stress tensor (A.1) reproduces
Eq.$\!$ (3.7) which contradicts the existence of solitons with finite
energy, the latter requirement being implicit in the transition from
Eq. (A.4) to Eq. (A.5) as well as in the original argument of Derrick [7].

In order to probe the existence of interesting static solutions with
infinite energy, we return to Eq.$\!$ (A.4) and take the limit 
$A\to\infty$ more carefully. For definiteness let us restrict our
attention to the axially symmetric configuration of Eq. (3.9), i.e.,
$\Theta =\theta (\rho)$ and $\Phi =\kappa (\phi +\phi_0)$ with
$\kappa =\pm 1$, which is inserted in the stress tensor (A.1) to
yield the reduced elements
$$\eqalign{\sigma_{11} & = -U\cos (2\phi)+V,\qquad
\sigma_{22}=U\cos (2\phi)+V,\cr
\noalign{\medskip}
\sigma_{12} & = -U\sin (2\phi)=\sigma_{21},\cr}\eqno \hbox{(A.7)}$$
where the functions
$$U={1\over 2}\left[\left({\partial\theta\over\partial\rho}\right)^2-
{\sin^2\theta\over\rho^2}\right] ,\qquad
V={1\over 2}\cos^2\theta ,\eqno \hbox{(A.8)}$$
depend only on the radial coordinate $\rho$. Furthermore the area A in 
Eq. (A.4) is taken to be a circle of radius $R$ and the Cartesian 
components of the line-element vector $d\nb s$ are expressed in terms 
of the polar coordinates as $ds_1=R\cos\phi\, d\phi$ and 
$ds_2=R\sin\phi\, d\phi$. Relations (A.4) applied for 
$\mu , \nu =1$ or 2 taken in all combinations reduce to the single 
virial relation
$$\int^R_0 2\pi\rho d\rho V(\rho)=\pi R^2[V(R)-U(R)],\eqno \hbox{(A.9)}$$
which must be satisfied for any radius $R$. Indeed, if we use the 
numerically calculated vortex profile $\theta =\theta (\rho)$ shown
in Fig.$\!$ 3 to calculate the functions $U=U(\rho)$ and $V=V(\rho)$
from Eqs.$\!$ (A.8), relation (A.9) is verified for all $R$. In particular,
the limit $R\to\infty$ yields a nonvanishing contribution in the 
right-hand side thanks to the centrifugal term in the function $U$
which leads to $U(R)\sim -1/2 R^2$ because
$\theta\to{\pi\over 2}$. The net result of this limit is 
$$\int^{\infty}_02\pi\rho d\rho \cos^2\theta =\pi ,\eqno \hbox{(A.10)}$$
which establishes the virial relation announced in Eq. (3.8).
This relation no longer contradicts the existence of nontrivial vortex
solutions but actually predicts analytically their anisotropy energy.
However the exchange energy is (logarithmically) divergent thanks
again to the centrifugal term.

\vfill
\eject
\bigskip
\noindent
{\ini REFERENCES}
\bigskip
\item{[1]\ }A.P. Malozemoff and J.C. Slonczewski, Magnetic domain 
walls in bubble materials (Wiley, New York, 1981).
\medskip
\item{[2]\ }V.G. Bar'yakhtar, M.V. Chetkin, B.A. Ivanov and S.N. Gadetskii,
Dynamics of topological magnetic solitons-experiment and theory
(Springer-Verlag, Berlin, 1994).
\medskip
\item{[3]\ }N. Papanicolaou, Phys. Rev. B {\bf 51}, 15062 (1995); and 
Dynamics of domain walls in weak ferromagnets (Crete preprint, 1996).
\medskip
\item{[4]\ }N.S. Manton, Phys. Lett. B {\bf 110}, 54 (1982); M. Atiyah
and N. Hitchin, The geometry and dynamics of magnetic monopoles
(Princeton University Press, Princeton, 1988); R.A. Leese, M. Peyrard
and W.J. Zakrzewski, Nonlinearity {\bf 3}, 773 (1990).
\medskip
\item{[5]\ }N. Papanicolaou and T.N. Tomaras, Nucl. Phys. B {\bf 360},
425 (1991).
\medskip
\item{[6]\ }S. Komineas and N. Papanicolaou, Topology and dynamics in 
ferromagnetic media, Physica D (in press).
\medskip
\item{[7]\ }G.H. Derrick, J. Math. Phys. {\bf 5}, 1252 (1964).
\medskip
\item{[8]\ }A.F. Andreev and V.I. Marchenko, Sov. Phys. Uspekhi
{\bf 23}, 21 (1980).
\medskip
\item{[9]\ }B.A. Ivanov and A.K. Kolezhuk, Phys. Rev. Lett. {\bf 74},
1859 (1995).
\medskip
\item{[10]\ }R.J. Donnely, Quantized vortices in helium II (Cambridge
University Press, Cambridge, 1991).
\medskip
\item{[11]\ }B.A. Ivanov, A.K. Kolezhuk and G.M. Wysin, Phys. Rev. Lett.
{\bf 76}, 511 (1996).
\medskip
\item{[12]\ }A.A. Belavin and A.M. Polyakov, JETP Lett. {\bf 22}, 245 (1975).
\medskip
\item{[13]\ }N. Papanicolaou and W.J. Zakrzewski, Physica D {\bf 80},
225 (1995); Phys. Lett. A {\bf 210}, 328 (1996).
\medskip
\item{[14]\ }R.S. Ward, Phys. Lett. B {\bf 158}, 424 (1985);
W.J. Zakrzewski, Nonlinearity {\bf 4}, 429 (1991);
R.A. Leese, Nucl. Phys. B {\bf 344}, 33 (1990).
\medskip
\item{[15]\ }P.J. Ruback, Nucl. Phys. B {\bf 296}, 669 (1988).
\medskip
\item{[16]\ }B.A. Ivanov and D.D. Sheka, Phys. Rev. Lett. {\bf 72},
404 (1994).
\medskip
\item{[17]\ }A.A. Thiele, Phys. Rev. Lett. {\bf 30}, 230 (1973);
J. Appl. Phys. {\bf 45}, 377 (1974).
\medskip
\item{[18]\ }N. Papanicolaou and T.N. Tomaras, Phys. Lett. A {\bf 179},
33 (1993).
\medskip
\item{[19]\ }G. Stratopoulos and T.N. Tomaras, 
Phys. Rev. B {\bf 54}, 12493 (1996).
\medskip
\item{[20]\ }G.K. Batchelor, An introduction to fluid dynamics (Cambridge
University Press, Cambridge, 1967).
\medskip
\item{[21]\ }A.R. V\"{o}lkel, G.M. Wysin, A.R. Bishop and F.G. Mertens,
Phys. Rev. B {\bf 44}, 10066 (1991).
\medskip
\item{[22]\ }A.R. Pereira and A.S.T. Pires, Phys. Rev. B {\bf 51},
996 (1995).

\vfill
\eject
\bigskip
\noindent
{\ini FIGURE CAPTIONS}
\bigskip
\noindent
{\bf Figure 1:\ }Illustration of the dimerization process for a finite
portion of the square lattice cut along the diagonals. Solid and open
circles denote the sites of the two intertwining sublattices.
\bigskip
\noindent
{\bf Figure 2:\ }Illustration of the tetramerization process of a finite
portion of the square lattice.
\bigskip
\noindent
{\bf Figure 3:\ }The numerically calculated profile of a static vortex
at weak anisotropy.
\bigskip
\noindent
{\bf Figure 4:\ }Spin vectors of a vortex and an antivortex projected
on the (12) plane. Spins are shown at every fourth site of the original 
lattice and distances on the grid are measured according to Eq. (2.24)
applied with $\varepsilon =0.1$.
\bigskip
\noindent
{\bf Figure 5:\ }Three characteristic snapshots of a head-on collision
of two like vortices originating at a relative distance $d=5$ on the 
$x$-axis with initial velocities $\nb v_1=(0.65,\,\, 0)$ and 
$\nb v_2=(-0.65,\,\, 0)$. After collision the two vortices scatter at 
$90^{\circ}$ and suffer an internal phase shift also equal to 
$90^{\circ}$. Vectors represent the projection of the field $\nb n$
on the (12) plane.
\bigskip
\noindent
{\bf Figure 6:\ }Three characteristic snapshots of the annihilation process
of a vortex and an antivortex that are initially at rest at a relative
distance $d=5$ on the $x$-axis. The vortex and the antivortex converge
toward each other and are eventually annihilated into spinwaves.
Vectors represent the projection of the field $\nb n$ on the (12) plane.
\bigskip
\noindent
{\bf Figure 7:\ }Level contours of the energy density corresponding to the
vortex-antivortex annihilation process shown in Fig.$\!$ 6. After collision
two distinct energy lumps emerge along the positive and negative
$y$-axis but eventually dissipate into spinwaves.
\bigskip
\noindent
{\bf Figure 8:\ }Schematic illustration of the ground state in the presence
of an external uniform magnetic field along the symmetry (third) axis.
The canting angle $\delta$ is given by Eq. (5.3).
\bigskip
\noindent
{\bf Figure 9:\ }Time evolution of a pair of like vortices in the presence
of a bias field $h=1$. The two vortices are initially at rest, at a 
relative distance $d=5$ on the $x$-axis, and subsequently orbit
around each other instead of drifting away. Together with the initial 
configuration of the field $\nb n$ projected on the (12) plane
(upper entry) we provide snapshots at instances when the pair had rotated
by $90^{\circ}$ (middle entry) and $180^{\circ}$ (lower entry). A 
$90^{\circ}$ rotation in real space is always followed by a $90^{\circ}$
internal phase shift.
\bigskip
\noindent
{\bf Figure 10:\ }Trajectories of the two rotating vortices of Fig. 9
obtained by tracking the points where $|n_3|=1$. The two vortices originated
at points A and B on the $x$-axis and the process was interrupted
after the pair had completed a $180^{\circ}$ rotation.
\bigskip
\noindent
{\bf Figure 11:\ }Evolution of the angular momentum of the vortex pair of 
Figs. 9 and 10 calculated from Eq. (5.54). The total angular momentum
$\ell$ (solid line) is fairly well conserved, except at late times when 
numerical instabilities develop on the finite lattice. The lower (upper)
dashed line depicts the nontrivial time dependence of the first (second)
term in Eq. (5.54).
\bigskip
\noindent
{\bf Figure 12:\ }Evolution of a vortex-antivortex pair in the presence of a 
bias field $h=1$. The vortex and the antivortex are initially at rest at the 
points $(2.4, -2.4)$ and $(-2.4, -2.4)$ of the $xy$-plane (upper
entry). Instead of converging toward each other and annihilating,
the pair moves in formation along the $y$-axis (Kelvin motion) as 
demonstrated by the snapshot of the lower entry taken at a later
instance $(\tau\approx 20)$.
\bigskip
\noindent
{\bf Figure 13:\ }Trajectories of the vortex and the antivortex of 
Fig.$\!$ 12
originating at the points $B=(2.4, -2.4)$ and $A=(-2.4, -2.4)$.
The actual trajectories (solid lines) were obtained by tracking the
points where $|n_3|=1$. The trajectories of the guiding centers
(dashed lines) were calculated from Eq. (5.55) restricted to the corresponding
half planes and are remarkably close to two parallel straight lines.
\bigskip
\noindent
{\bf Figure 14:\ }Evolution of the linear momentum of the 
vortex-antivortex pair of Figs.$\!$ 12 and 13 calculated from Eq.$\!$ (5.53).
(We display only the $y$-component because the $x$-component vanishes.)
The total linear momentum (solid line) is fairly well conserved,
whereas the lower (upper) dashed line depicts the nontrivial time
dependence of the first (second) term of Eq. (5.53).

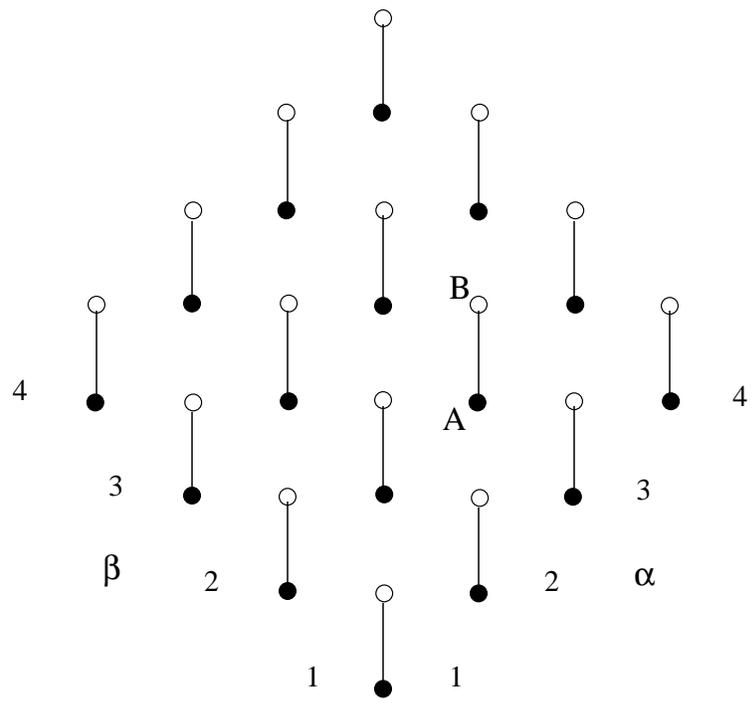
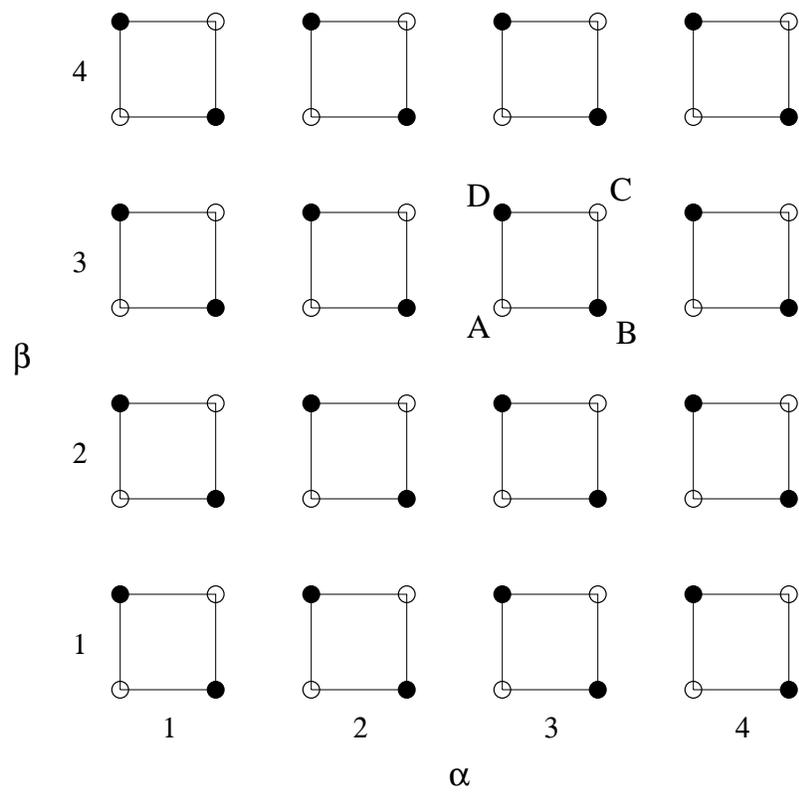
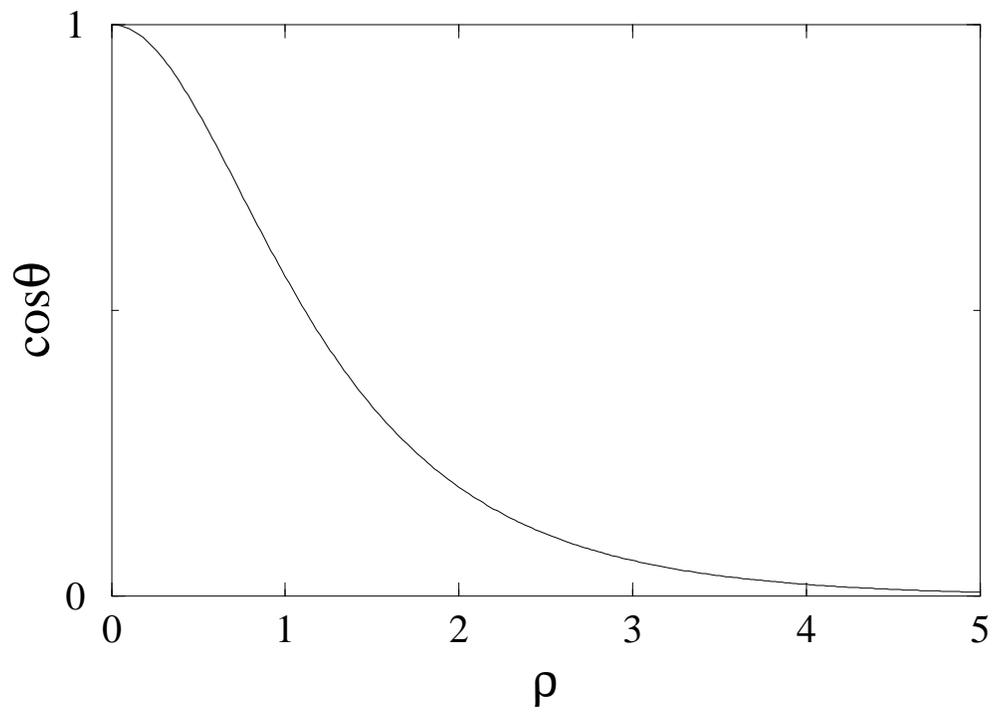
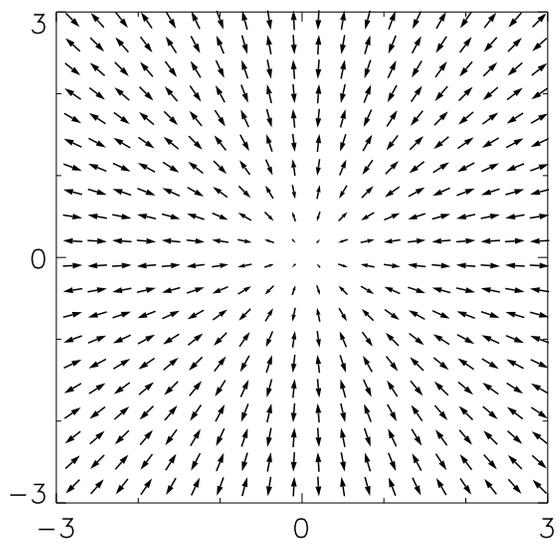
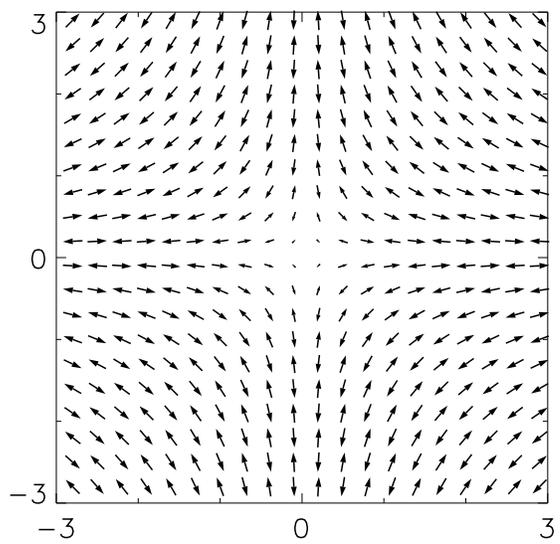
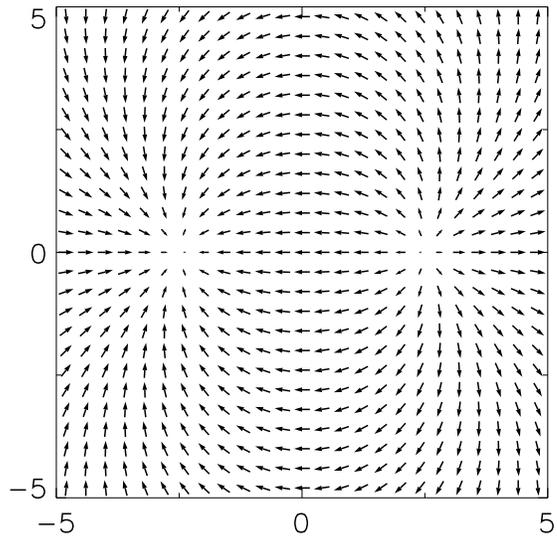
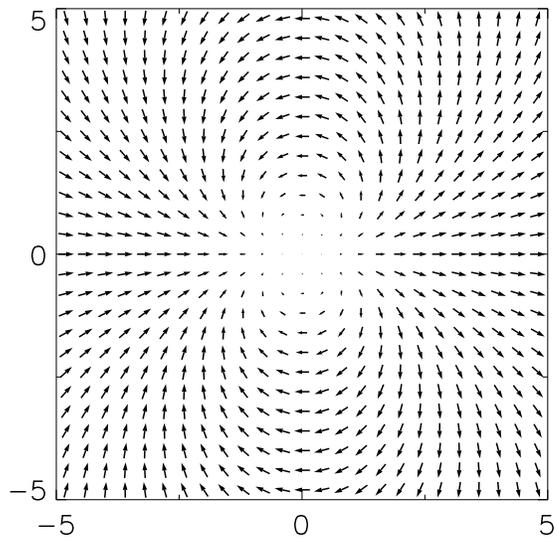
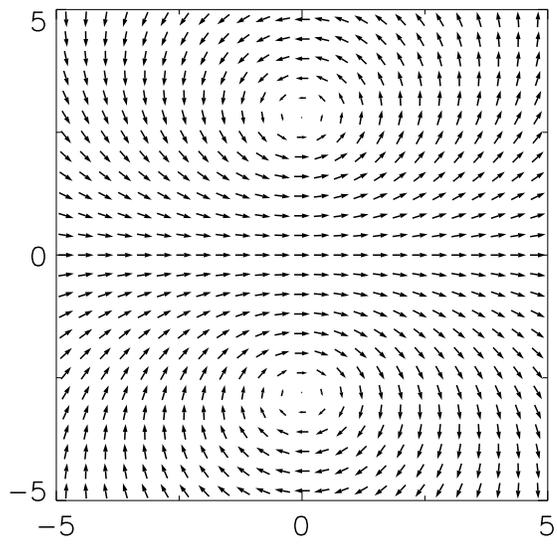
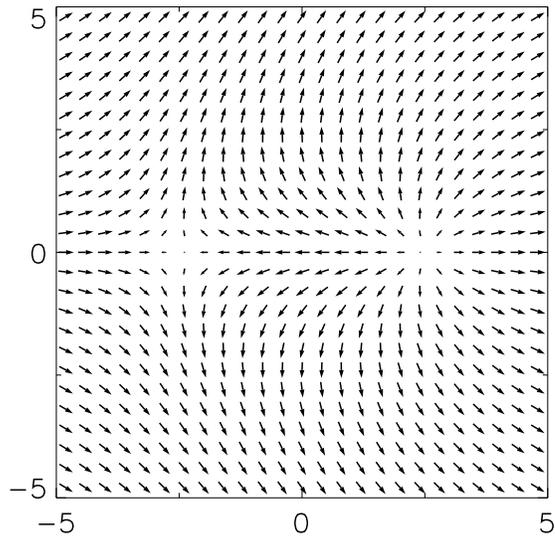
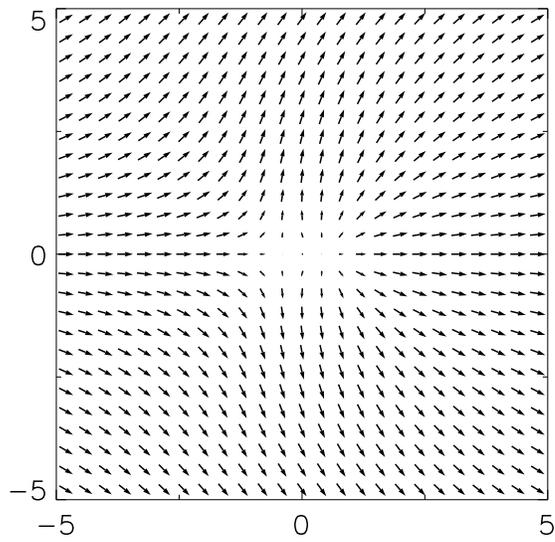
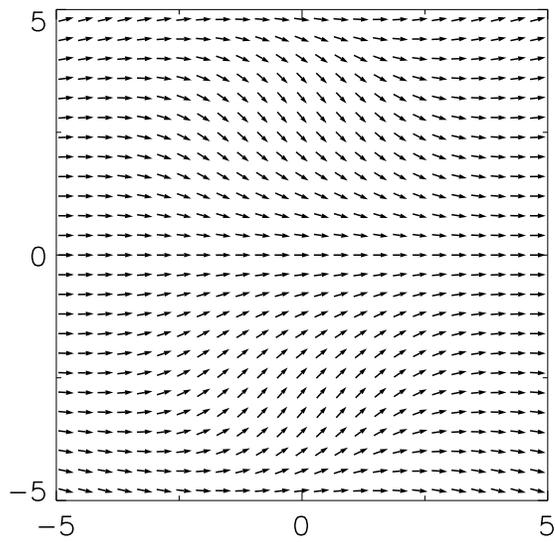
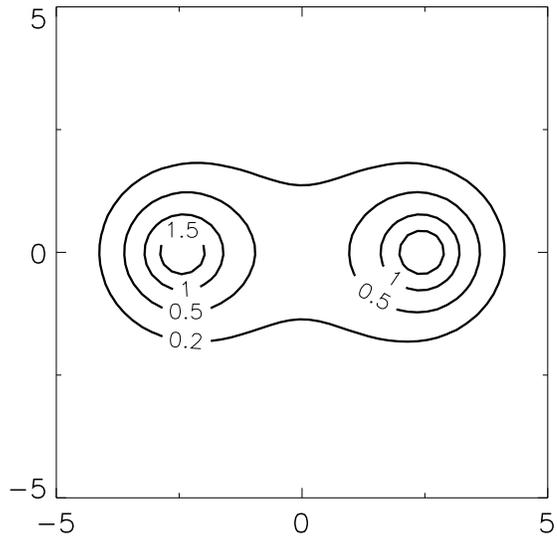
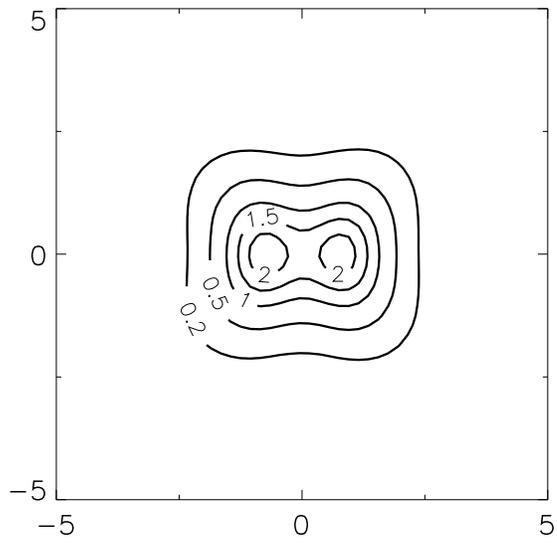
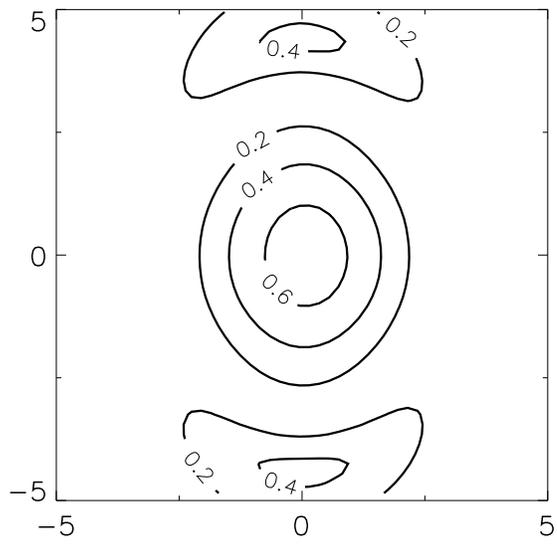
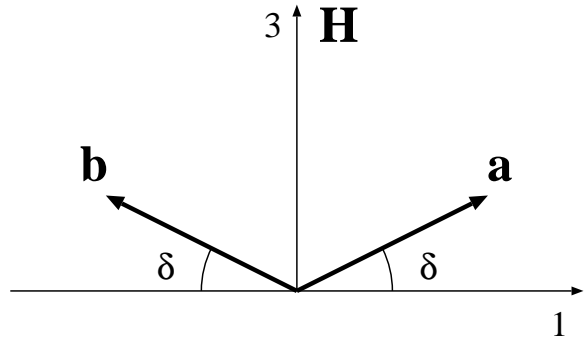
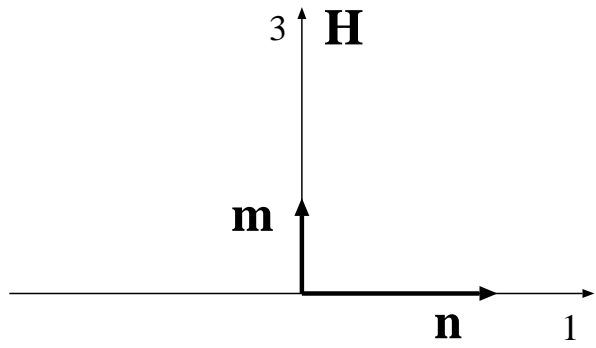
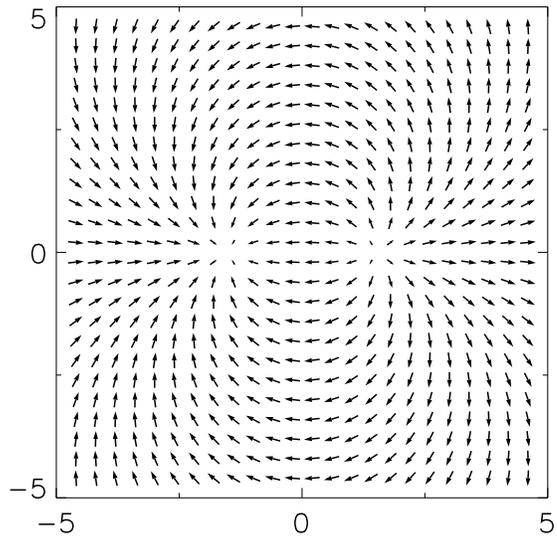
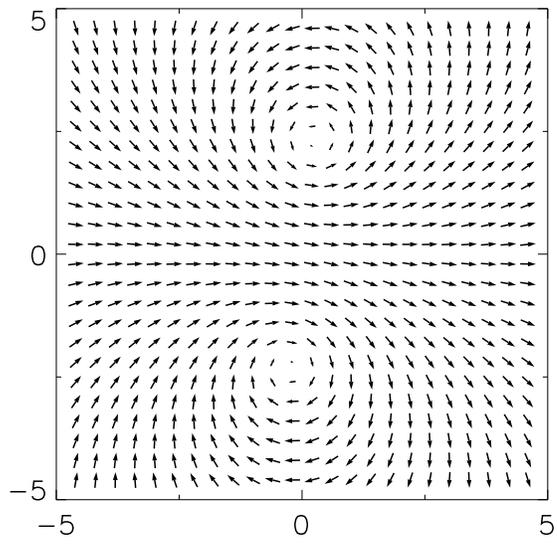
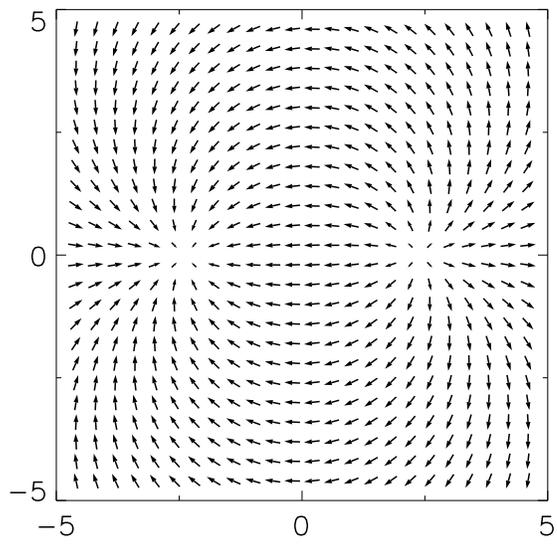
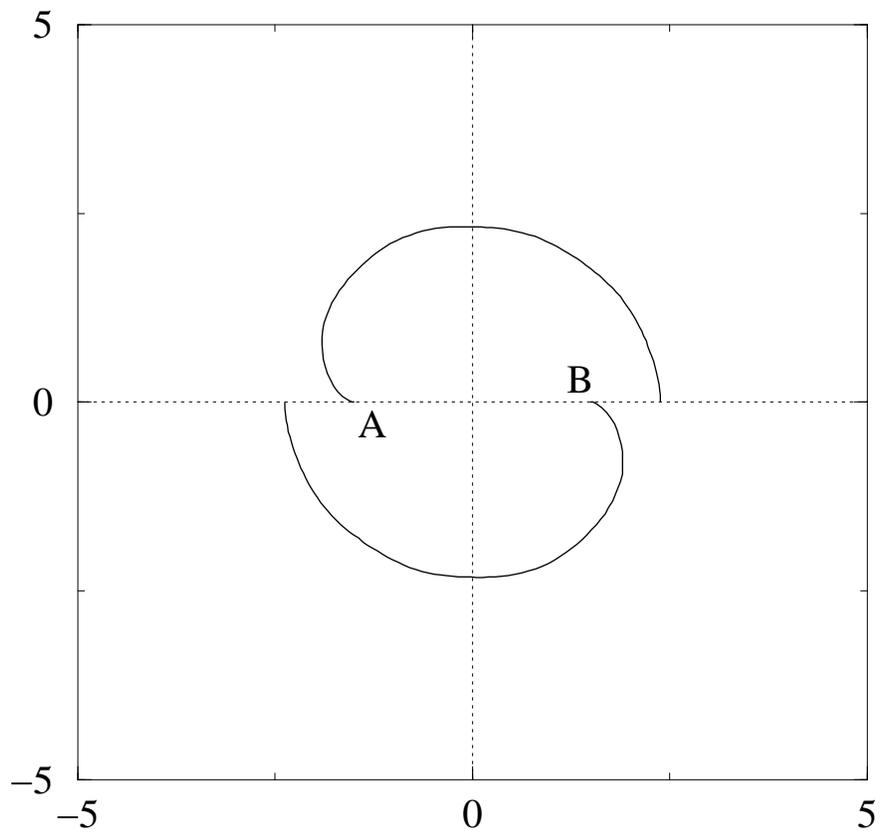
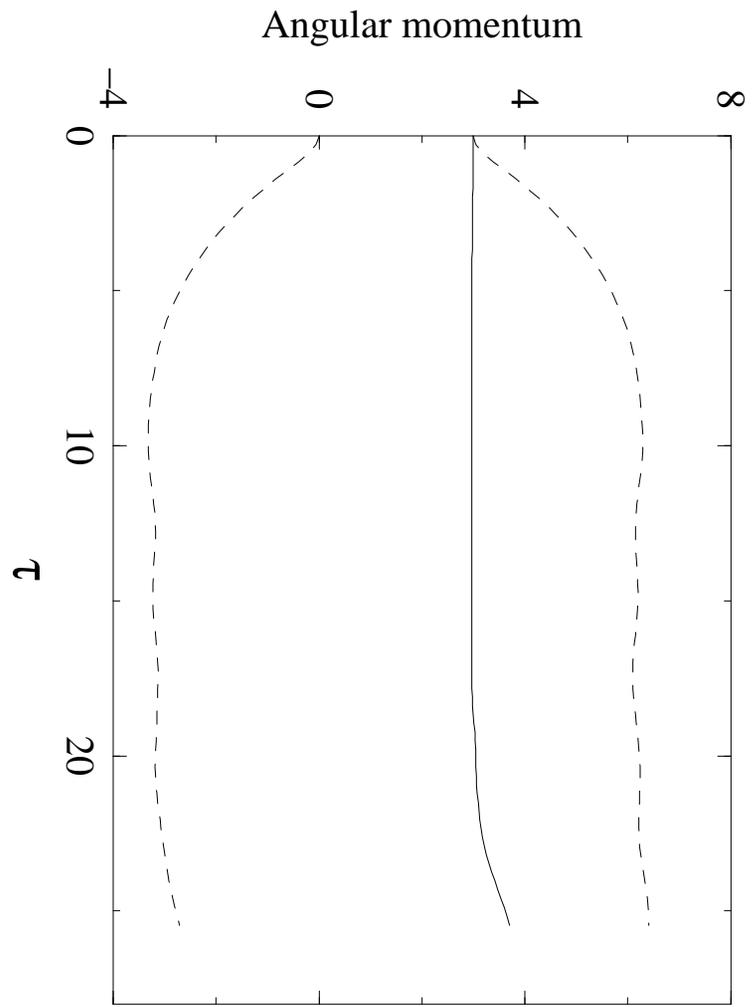
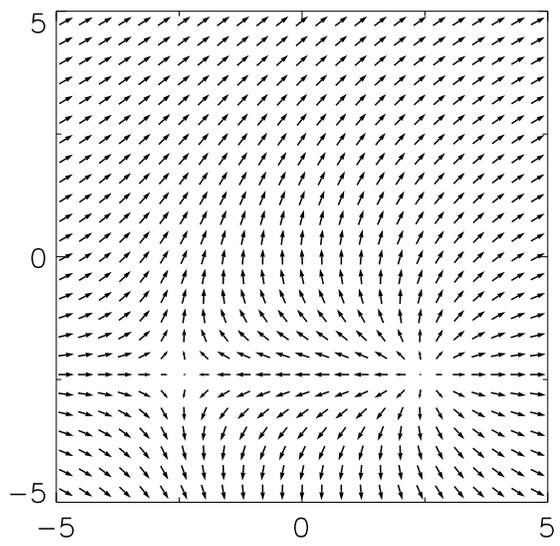
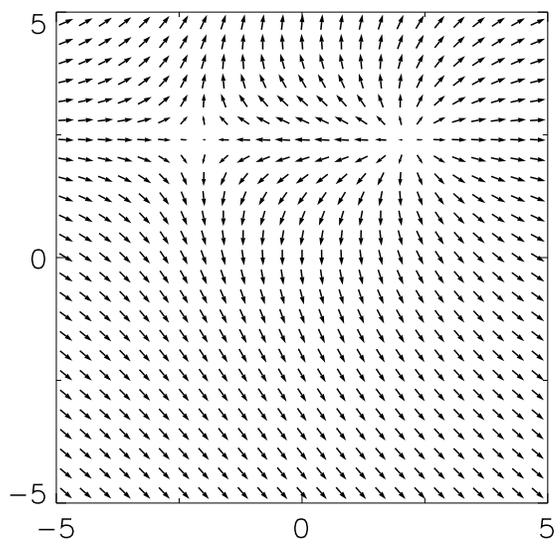
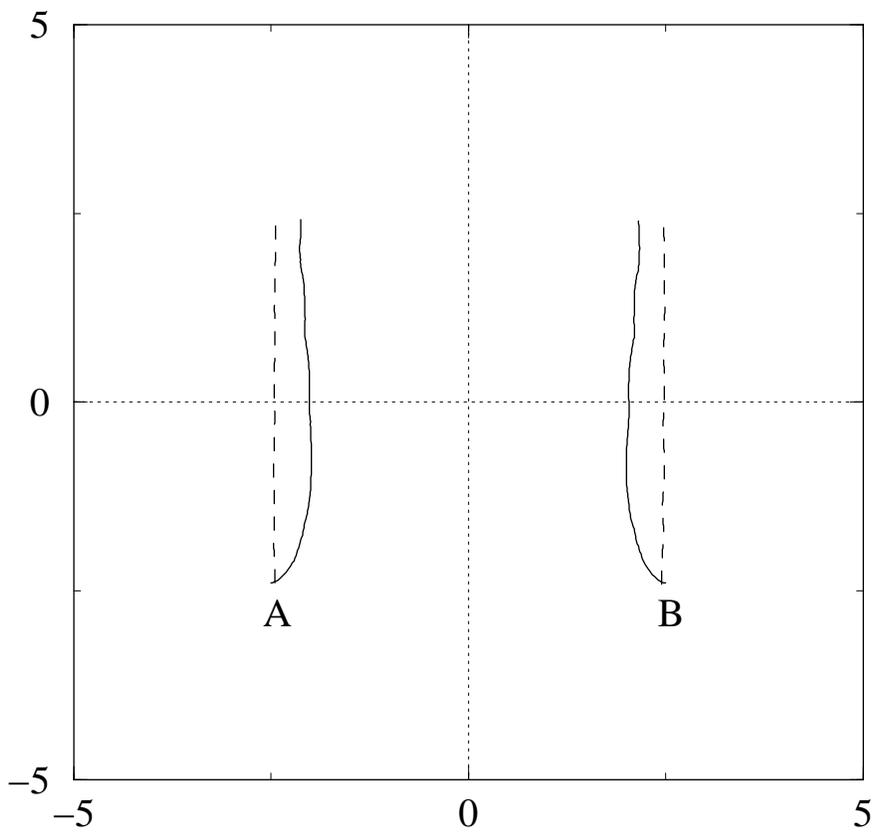
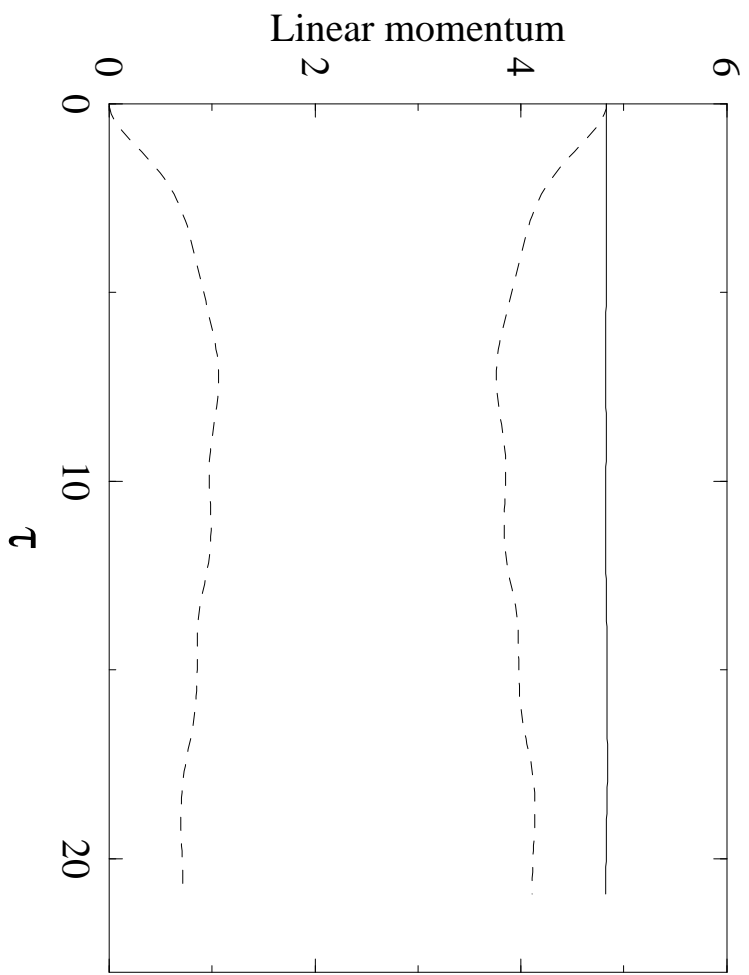
\end